\documentclass[aip,pop,reprint,numerical]{revtex4-1}
\usepackage[utf8]{inputenc}
\usepackage{amssymb}
\usepackage{amsmath}
\usepackage{bm}
\usepackage{graphicx}
\usepackage[space]{grffile}
\usepackage{xspace}
\usepackage{color}
\usepackage[center]{subfigure}
\usepackage{booktabs}
\usepackage{wasysym}
\usepackage{multirow}
\usepackage{aas_macros}

\newcommand{\Ms}{\mathrm{M_s}}
\newcommand{\Ma}{\mathrm{M_a}}
\renewcommand{\Re}{\mathrm{Re}}
\newcommand{\Rm}{\mathrm{Rm}}

\newcommand{\apost}{\textit{a posteriori}\xspace}
\newcommand{\aprio}{\textit{a priori}\xspace}
\newcommand{\pd}[2]{\frac{\partial #1}{\partial #2}}
\newcommand{\flt}[1]{\overline{#1}}
\newcommand{\fav}[1]{\widetilde{#1}}

\newcommand{\Enzo}{\textsc{Enzo}\xspace}
\newcommand{\FLASH}{\textsc{FLASHv4}\xspace}
\newcommand{\VEC}[1]{\boldsymbol{ #1}}
\newcommand{\bottomrulea}[1]{\left( #1 \right)}
\newcommand{\grad}[1]{\nabla \bottomrulea{#1}}
\newcommand{\curl}[1]{\nabla \times #1}
	
\renewcommand{\div}[1]{\nabla \cdot \bottomrulea{#1}}

\newcommand{\EMFdata}{{\bm{\mathcal{E}}}}
\newcommand{\rres}{\overline{\rho}}
\newcommand{\jres}{\overline{\mathbf{J}}}
\newcommand{\Jflt}{\jres}
\newcommand{\wres}{\tilde{\mathbf{\Omega}}}
\newcommand{\Wflt}{\wres}
\newcommand{\pres}{\overline{P}}

\newcommand{\EMF}{\widehat{\bm{\mathcal{E}}}}
\newcommand{\tuijdata}[1][]{\tau_{ij}^{\mathrm{u} #1}}
\newcommand{\tbijdata}[1][]{\tau_{ij}^{\mathrm{b} #1}}
\newcommand{\tudata}[1][]{\tau^{\mathrm{u} #1}}
\newcommand{\tbdata}[1][]{\tau^{\mathrm{b} #1}}
\newcommand{\Ekinsgsdata}{E^{\mathrm{u}}_{\mathrm{sgs}}}
\newcommand{\Emagsgsdata}{E^{\mathrm{b}}_{\mathrm{sgs}}}
\newcommand{\Esgsdata}{E_{\mathrm{sgs}}}
\newcommand{\Wsgsdata}{W_{\mathrm{sgs}}}
\newcommand{\EkinsgsS}{\widehat{E}^{\mathrm{u},\mathcal{S}}_{\mathrm{sgs}}}
\newcommand{\EkinsgsSStar}{\widehat{E}^{\mathrm{u},\mathcal{S}^*}_{\mathrm{sgs}}}

\newcommand{\EmagsgsM}{\widehat{E}^{\mathrm{b},\mathcal{M}}_{\mathrm{sgs}}}
\newcommand{\EmagsgsJ}{\widehat{E}^{\mathrm{b},J}_{\mathrm{sgs}}}
\newcommand{\EkinsgsSS}{\widehat{E}^{\mathrm{u},\mathrm{SS}}_\mathrm{sgs}}
\newcommand{\EkinsgsNL}{\widehat{E}^{\mathrm{u},\mathrm{NL}}_\mathrm{sgs}}
\newcommand{\EmagsgsSS}{\widehat{E}^{\mathrm{b},\mathrm{SS}}_\mathrm{sgs}}
\newcommand{\EmagsgsNL}{\widehat{E}^{\mathrm{b},\mathrm{NL}}_\mathrm{sgs}}

\newcommand{\nuk}{\nu^\mathrm{u}}

\newcommand{\nub}{\nu^\mathrm{b}}
\newcommand{\EVconst}{\mathrm{EV}^\mathrm{const}}
\newcommand{\EVSM}{\mathrm{EV}^\mathcal{SM}}
\newcommand{\EVW}{\mathrm{EV}^W}
\newcommand{\EVSStar}{\mathrm{EV}^\mathcal{S^*}}
\newcommand{\EVE}{\mathrm{EV}^{E}}
\newcommand{\ED}{\mathrm{ED}}
\newcommand{\SSu}{\mathrm{SS}^\mathrm{u}}
\newcommand{\SSid}{\mathrm{SS}}
\newcommand{\NLu}{\mathrm{NL^u}}
\newcommand{\NL}{\mathrm{NL}}
\newcommand{\NLuSStar}{\mathrm{NL}^{\mathrm{u,\mathcal{S^*}}}}
\newcommand{\NLuE}{\mathrm{NL}^{\mathrm{u,E}}}
\newcommand{\EDconst}{\mathrm{ED}^\mathrm{const}}                   
\newcommand{\EDW}{\mathrm{ED}^W}
\newcommand{\EDM}{\mathrm{ED}^\mathcal{M}}
\newcommand{\EDE}{\mathrm{ED}^{E}}
\newcommand{\SSb}{\mathrm{SS}^\mathrm{b}}
\newcommand{\NLb}{\mathrm{NL^b}}
\newcommand{\NLbM}{\mathrm{NL}^{\mathrm{b,\mathcal{M}}}}
\newcommand{\NLbE}{\mathrm{NL}^{\mathrm{b,E}}}
\newcommand{\ERconst}{\mathrm{ER}^\mathrm{const}}
\newcommand{\ERSM}{\mathrm{ER}^\mathcal{SM}}
\newcommand{\ERW}{\mathrm{ER}^W}
\newcommand{\ERSpM}{\mathrm{ER}^\mathcal{S+M}}
\newcommand{\ERE}{\mathrm{ER}^{E}}
\newcommand{\abg}{\alpha\text{-}\beta\text{-}\gamma}
\newcommand{\SSemf}{\mathrm{SS}^\mathcal{E}}
\newcommand{\NLemf}{\mathrm{NL}^\mathcal{E}}                   
\newcommand{\NLemfcompr}{\mathrm{NL}^{\mathcal{E},\rho}}
\newcommand{\tu}[1][]{\widehat{\tau}_{ij}^{\mathrm{u} #1}}
\newcommand{\tb}[1][]{\widehat{\tau}_{ij}^{\mathrm{b} #1}}
\newcommand{\tunoij}[1][]{\widehat{\tau}^{\mathrm{u} #1}}

\newcommand{\Sflt}[1][]{\fav{\mathcal{S}}}
\newcommand{\Mflt}[1][]{\flt{\mathcal{M}}}

\newcommand{\pb}{\beta_{\mathrm{p}}}

\newcommand{\abs}[1]{\left | #1 \right |}
\newcommand{\SE}{\Sigma^E}
\newcommand{\SW}{\Sigma^W}
\newcommand{\FE}{\mathcal{F}^E}
\newcommand{\FW}{\mathcal{F}^W}
\newcommand{\bra}[1]{\left( #1 \right)}
\DeclareMathOperator{\sgn}{sgn}
\DeclareMathOperator{\tr}{tr}

\makeatletter
\def\widebreve#1{\mathop{\vbox{\m@th\ialign{##\crcr\noalign{\kern3\p@}%
      \brevefill\crcr\noalign{\kern3\p@\nointerlineskip}%
      $\hfil\displaystyle{#1}\hfil$\crcr}}}\limits}

\def\brevefill{$\m@th \setbox\z@\hbox{$\braceld$}%
  \bracelu\leaders\vrule \@height\ht\z@ \@depth\z@\hfill\braceru$}
\makeatletter

\begin{document}

\title[A priori comparison of MHD SGS closures]{A nonlinear structural subgrid-scale closure for compressible MHD Part II: a priori comparison on turbulence simulation data}
\author{Philipp Grete}
\email{grete@mps.mpg.de}
\affiliation{Max-Planck-Institut für Sonnensystemforschung, Justus-von-Liebig-Weg 3, D-37077 Göttingen, Germany}
\affiliation{Institut für Astrophysik, Universität Göttingen, Friedrich-Hund-Platz 1, D-37077 Göttingen, Germany}
\author{Dimitar G Vlaykov}
\affiliation{Max-Planck-Institut für Dynamik und Selbstorganisation, Am Faßberg 17, D-37077 Göttingen, Germany}
\affiliation{Institut für Astrophysik, Universität Göttingen, Friedrich-Hund-Platz 1, D-37077 Göttingen, Germany}
\author{Wolfram Schmidt}
\affiliation{Hamburger Sternwarte, Universität Hamburg, Gojenbergsweg 112, D-21029 Hamburg, Germany}
\author{Dominik R G Schleicher}
\affiliation{Departamento de Astronomía, Facultad Ciencias Físicas y Matemáticas, Universidad de Concepción,  Av. Esteban Iturra s/n Barrio Universitario, Casilla 160-C, Chile}

\begin{abstract}
Even though compressible plasma turbulence is encountered in many astrophysical phenomena, its effect is often not well understood.
Furthermore, direct numerical simulations are typically not able to reach the extreme parameters of these processes.
For this reason, large-eddy simulations (LES), which only simulate large and intermediate scales directly, are employed.
The smallest, unresolved scales and the interactions between small and large scales are introduced by means
of a subgrid-scale (SGS) model.
We propose and verify a new set of nonlinear SGS closures for future application as an SGS model 
in LES of compressible magnetohydrodynamics (MHD).
We use 15 simulations (without explicit SGS model) of forced, isotropic, homogeneous turbulence with 
varying sonic Mach number $\Ms = 0.2$ to $20$
as reference data for the most extensive \aprio tests performed so far in literature.
In these tests we explicitly filter the reference data and compare the performance of the new closures against the 
most widely tested closures.
These include eddy-viscosity and scale-similarity type closures with different normalizations.
Performance indicators are correlations with the turbulent energy and cross-helicty flux, 
the average SGS dissipation, the topological structure and the ability to reproduce the
correct magnitude and direction of the SGS vectors.
We find that only the new nonlinear closures exhibit consistently high correlations (median value \textgreater$0.8$) 
with the data over the entire parameter space
and outperform the other closures in all tests.
Moreover, we show that these results are independent of resolution and chosen filter scale.
Additionally, the new closures are effectively coefficient-free with a deviation of less than $20\%$. 

\end{abstract}
\pacs{52.35.Ra, 52.65.Kj, 52.30.Cv, 47.27.em}
\date{\today}

\maketitle

\section{Introduction}
Turbulence and in particular plasma turbulence is still one of the least understood
phenomena in classical physics today.
Even though there are advances in theory many processes cannot be 
fully explained yet due to their strong nonlinearity.
These cover many different scales and 
include experiments on Earth \cite{2014PhPl...21a3505C}
as well as a wide variety of processes (e.g. 
magnetic reconnection \cite{2015ApJ...806L..12O}
and turbulent dynamos\cite{PhysRevE.92.023010}) and
astrophysical phenomena such as 
stellar winds \cite{Goldstein1995} and
magnetized accretion disks \cite{RevModPhys.70.1}.
Compressibility also plays an important role in astrophysical plasmas
and increases the complexity even further.

In addition to theory, experiments and observations, 
numerical simulations are a useful tool to understand turbulence.
However, the level of detail is restricted by the 
available computing power and realistic (physical) dynamical ranges are 
usually not covered.
Fortunately, this problem can be improved with the help of 
large eddy simulations (LES)\cite{lrca-2015-2,Miesch2015}.
This approach simulates only the largest and intermediate scales directly.
The smallest scales, which are below the resolution limit, i.e. below the grid scale,
are introduced by means of a subgrid-scale (SGS) model.
Formally, the procedure involves the convolution of the primary equations
with a filter kernel $G$.
For a static, homogeneous and isotropic filter 
the compressible magnetohydrodynamics (MHD) equations 
under boundary conditions read\cite{Chernyshov2014,Vlaykov15}
\begin{eqnarray}
\label{eq:masscons}
  \pd{\rres}{t}+\div {\rres \fav{\VEC{u}}} = 0,\\
\label{eq:momcons}
  \begin{split}
  \pd{\flt{\rho} \fav{\VEC{u}}}{t} 
  + \div{\flt{\rho} \fav{\VEC{u}} \otimes \fav{\VEC{u}} 
    - \flt{\VEC{B}} \otimes \flt{\VEC{B}}} 
     + \grad{\pres + \frac{\flt{B}^2}{2}}  \\*
    = \nabla \cdot \bra{2\nu\flt{\rho} \Sflt^{*}}
    - \nabla \cdot \tau, 
    \end{split} \\
\label{eq:fluxcons}
    \pd{\flt{\VEC{B}}}{t} - \curl\bra{\fav{\VEC{u}} \times \flt{\VEC{B}}} +
    \eta \nabla^2 \flt{\VEC{B}}=
    \curl{\EMFdata}.
\end{eqnarray}
Filtering is denoted by $\flt{\Box}$ and mass-weighted filtering
\cite{1983PhFl...26.2851F} is denoted by $\fav{\Box} = \flt{\rho\Box}/\flt{\Box}$.
Thus, $\flt{\rho}$, $\fav{\VEC{u}}$, $\flt{\VEC{B}}$ (incorporating $1/\sqrt{4 \pi}$)
and $\flt{P}$ are the filtered density, velocity, magnetic field and thermal pressure,
respectively.
In the context of LES filtered quantities are considered resolved and therefore 
accessible in the simulation.
Non-ideal effects are included via  resistivity $\eta$ and
kinematic viscosity $\nu$ with traceless kinetic 
rate-of-strain tensor 
$\Sflt^{*}_{ij}~=~1/2 \bra{\fav{u}_{i,j} + \fav{u}_{j,i}} - 1/3 \delta_{ij}\fav{u}_{k,k}$.
Here, $\Box_{i,j}$ designates the $j$-th partial derivative of component $i$,
a star $\Box_{ij}^{*}$ 
indicates the traceless, deviatoric
part of a tensor, and
Einstein summation convention applies with the Kronecker delta $\delta_{ij}$.
Two new terms enter the equations \eqref{eq:momcons} and \eqref{eq:fluxcons}.
The first term, modified from its hydrodynamical form,
is the turbulent stress tensor 
\begin{eqnarray}
  \label{eq:tau_def}\tau_{ij} =  \tuijdata - \tbijdata +
  \bra{\flt{B^2} - \flt{B}^2}\frac{\delta_{ij}}{2} \quad \text{with} 
  \qquad \qquad \\
  \tuijdata \equiv \flt{\rho} \bra{\fav{u_i u_j} - \fav{u}_i \fav{u}_j}
  \quad \text{and} \quad
  \tbijdata \equiv \bra{\flt{B_i B_j} - \flt{B}_i~\flt{B}_j} \quad \label{eq:tudata}
\end{eqnarray}
which consists of the turbulent (or SGS) magnetic pressure 
(last term in \eqref{eq:tau_def}), 
the SGS Reynolds stress $\tuijdata$ and the SGS Maxwell Stress $\tbijdata$.
The second term is the turbulent electromotive force (EMF)
\begin{eqnarray}
  \EMFdata = \flt{\VEC{u}\times\VEC{B}} - \fav{\VEC{u}} \times \flt{\VEC{B}}\label{eq:EMFdata}
\end{eqnarray}
in the induction equation. 
Both terms are \aprio unknown as only filtered primary
quantities are accessible in LES (e.g. $\fav{\VEC{u}}$) but 
no mixed terms (e.g. $\fav{u_i~u_j}$).
Moreover, the total filtered energy density 
\begin{eqnarray}
	\label{eq:SGSen}
	\flt{E} = \underbrace{\frac{1}{2} \flt{\rho} \fav{u}^2 + \frac{1}{2}\flt{B}^2}_
	{\mathrm{(resolved)}}
	+ \underbrace{\frac{1}{2}\flt{\rho} \bra{\fav{u^2} - \fav{u}^2} + \frac{1}{2} \bra{\flt{B^2} - \flt{B}^2}}_
	{= \Ekinsgsdata + \Emagsgsdata \equiv \Esgsdata \mathrm{(unresolved)}}
\end{eqnarray}
contains unclosed terms as well, namely the kinetic SGS energy $\Ekinsgsdata$ and 
magnetic SGS energy $\Emagsgsdata$.
These terms are 
given by the isotropic parts of the turbulent stress tensors
\begin{eqnarray}\label{eq:EsgsTau}
\frac{1}{2} \tau^\Box_{kk} = \Esgsdata^\Box \;.
\end{eqnarray}
Similarly, the filtering procedure applies to other quantities such as the 
cross-helicity $W$ -- a measure of the alignment between velocity 
and magnetic field.
The resulting SGS cross-helicity $\Wsgsdata$ is given by
\begin{eqnarray}
	\label{eq:Wsgs}
\Wsgsdata = \flt{\VEC{u}\cdot \VEC{B}} - \fav{\VEC{u}}\cdot\flt{\VEC{B}}
\end{eqnarray}
It encodes not only the alignment between unresolved fields, 
but also between resolved and unresolved ones.

On the one hand, there has been a lot of research in the realm of 
(incompressible) hydrodynamics\cite{Sagaut2006} with successful
applications to atmospheric boundary layers\cite{LuPorte-Agel10} and
turbulent mixing \cite{FLM:13495,Balarac2013}, as well as
astrophysical application\cite{lrca-2015-2} in different subjects such as 
isolated disc galaxies\cite{Braun21082014} or 
the formation of supermassive black holes\cite{2013MNRAS.436.2989L}.
On the other hand, results for MHD are still scarce and limited to \apost 
application of (decaying) turbulent boxes\cite{Miki2008} in either 2D\cite{1994PhPl....1.3016T},
or in the incompressible case\cite{Mueller2002,Yokoi2013}, or by neglecting 
terms such as turbulent magnetic pressure\cite{Chernyshov2014}.
However, the \aprio validation of these closures is still outstanding, apart
from a single incompressible dataset for the EMF\cite{Balarac2010}.
For this reason, we here expand our first investigation of nonlinear
closures\cite{Grete2015} with additional closures from the literature, and over
a more extended set of parameters and test cases.
We have identified several closure strategies developed in the literature and
evaluate the three major ones: 
eddy-viscosity, which is typically purely dissipative, 
scale-similarity, which is based on the self-similar properties of turbulence,
and deconvolution closures, which are fundamentally nonlinear based on 
approximate inverses of the filtering operator.
All closures, including the new nonlinear closures,
are briefly presented in the next section.
A detailed derivation and formal analysis of the new closures
is described in our accompanying paper\cite{Vlaykov2016a}.
In section~\ref{sec:verification} we describe our test setup and the process
of \aprio testing for several reference quantities.
The results  are then illustrated in section~\ref{sec:results} and 
include a wide variety of functional and structural tests.
Finally, in section~\ref{sec:conclusions} we conclude with an overall
comparison of the presented closures.

\section{Closures}
\label{sec:closures}
The following independent terms require closures:
the SGS Reynolds stress
$\tudata$, the SGS Maxwell stress $\tbdata$ and the electromotive force $\EMFdata$.
In the following, we briefly present three general closure strategies (eddy-viscosity, 
scale-similarity and nonlinear) and possible variations with respect to 
normalization.
Each closure strategy is based on a certain idea that naturally transfers
to closures of all unknown terms.
We identify closures by two uppercase roman letters 
(with normalizations in superscript), and closure expressions in formulas 
are denoted by a hat $\widehat{\Box}$.

\paragraph{The eddy-dissipation} 
family is the most well-established type of closure originating from 
the Smagorinsky eddy-viscosity\cite{Smagorinsky1963}
going back several decades.
In general, the modeled effects are purely dissipative in nature and resemble
existing terms, e.g. the Reynolds stress~\eqref{eq:tu} has the same functional form as 
the microscopic dissipation in the momentum equation, c.f. the right hand side of 
\eqref{eq:momcons}.
The same is true for the EMF~\eqref{eq:ER} and Ohmic dissipation 
in the induction equation.
An eddy-diffusivity based closure for the Maxwell stress
has been proposed\cite{Miki2008} analogous to eddy-viscosity.
The resulting closures are
\begin{eqnarray}
	\label{eq:tu}
	\mathrm{EV:} \qquad
	\tu[*] &=& -2  \flt{\rho} \nuk \mathcal{\fav{S}}^*_{ij} \;,\\
	\label{eq:tb}
	\mathrm{ED:} \qquad
	\tb[*] &=& -2  \nub \Mflt_{ij} \;, \\
	\label{eq:ER}
	\mathrm{ER:} \qquad \;
	\EMF &=& -  \eta_t \Jflt \;,
\end{eqnarray}
with eddy-viscosity ($\mathrm{EV}$)$ \nuk$, 
diffusivity ($\mathrm{ED}$) $\nub$, 
resistivity ($\mathrm{ER}$) $\eta_t$, 
and resolved current $\Jflt = \nabla \times \flt{\VEC{B}}$.
The kinetic rate-of-strain tensor 
$\mathcal{\fav{S}}^*_{ij}$
and magnetic rate-of-strain tensor 
$\Mflt_{ij} = 1/2 \bra{\flt{B}_{i,j} + \flt{B}_{j,i}}$ 
are by construction deviatoric and so are the closures
\eqref{eq:tu} and \eqref{eq:tb}.
The remaining isotropic parts are closed by means of
SGS energy closures
\begin{eqnarray}
\label{eq:EnClosures}
\EmagsgsM = C_1 \Delta^2 |\Mflt|^2 \quad 
\text{and} \quad
\EkinsgsSStar = C_2 \Delta^2 \flt{\rho} |\Sflt^*|^2 , \;
\end{eqnarray}
which can be derived from \eqref{eq:tu} and \eqref{eq:tb} building upon
the realizability of
$\tu$ and $\tb$ for a positive filter kernel\cite{Vreman1994,Grete2015}.
The free coefficients $C_\Box$ appear independently in every closure (including
all following ones) and are typically dimensionless.
One goal of \aprio testing is the determination of the coefficient values
as described in subsection \ref{ss:refQuant}.

In addition to the realizability ansatz, 
the isotropic parts can be closed under the assumption of 
local equilibrium between production and dissipation in the 
SGS energy evolution equations\cite{Vlaykov15} resulting in
\begin{eqnarray}
\label{eq:EnClosuresSmag}
\EmagsgsJ = C_3 \Delta^2 |\flt{\VEC{J}}|^2 \quad 
\text{and} \quad
\EkinsgsS = C_4 \Delta^2 \flt{\rho} |\Sflt|^2 \;.
\end{eqnarray}

Furthermore, several normalizations (or scalings)
have been developed to control the strength
of the deviatoric closures based on different arguments.
In this paper, we test the most often used ones, i.e. constant
scaling, scaling by SGS energy, and scaling by 
the interaction between the velocity and the magnetic field.
Constant scaling is given by
\begin{eqnarray}
\EVconst: \qquad \nuk &=& C_5 \Delta^{4/3} \;,  \\
\EDconst: \qquad \nub &=& C_6 \Delta^{4/3} \;,  \\
\ERconst: \qquad \eta_t &=& C_7 \Delta^{4/3} \;, 
\end{eqnarray}
motivated by dimensional analysis under Kolmogorov scaling\cite{Agullo2001}.
These closures neglect any local variability of the eddy-viscosity, 
diffusivity and resistivity.
In contrast to this, SGS energies, as a local measure of unresolved turbulence,
can be used as a proxy to obtain spatially varying closures
\begin{eqnarray}
\EVE: \qquad
\nuk &=& C_{13} \Delta \sqrt{\Ekinsgsdata/\flt{\rho}}
\;, \\
\EDE: \qquad
\nub &=& C_{14} \Delta \sqrt{\Emagsgsdata}
\;, \\
\ERE: \qquad
\eta_t &=& C_{15} \Delta \sqrt{\bra{\Ekinsgsdata + \Emagsgsdata}/\flt{\rho}} 
\;.
\end{eqnarray}
However, the exact values for the energies $\Ekinsgsdata$ and $\Emagsgsdata$
\eqref{eq:SGSen} are unknown.
Thus, the energy closure expressions \eqref{eq:EnClosures} can be used to
formulate complete closures\cite{1994PhPl....1.3016T} based only on known fields
\begin{eqnarray}
\EVSStar: \qquad
\nuk &=& C_{16} \Delta \sqrt{\EkinsgsSStar/\flt{\rho}} 
\;, \\
\EDM: \qquad
\nub &=& C_{17} \Delta \sqrt{\EmagsgsM} 
\;, \\
\ERSpM: \qquad
\eta_t &=& C_{18} \Delta \sqrt{\bra{\EkinsgsSStar + \EmagsgsM}/\flt{\rho}} \;.
\end{eqnarray}
Another possibility to include local variability is via the 
interactions between velocity and magnetic field.
Here, the SGS cross-helicity \eqref{eq:Wsgs} 
serves as a proxy in the closures
\begin{eqnarray}
\EVW: \qquad
\nuk &=& C_{10}\Delta \rho^{-1/4} \sqrt{|\Wsgsdata|} 
\;, \\
\EDW: \qquad
\nub &=& C_{11} t_t \Wsgsdata 
\;, \\
\ERW: \qquad
\eta_t &=& C_{12} t_t \sqrt{\flt{\rho}} \Wsgsdata
\;.
\end{eqnarray}
with a turbulent time scale $t_t = \Delta \sqrt{\flt{\rho}/\Esgsdata}$.
Again, an alternative formulation has been proposed\cite{Mueller2002}
\begin{eqnarray}
\EVSM: \quad
\nuk &=& C_8 \Delta^2 \rho^{-1/4} \sqrt{|2\Sflt_{ij}\Mflt_{ij}|}
 \\
\ERSM: \quad
\eta_t &=& C_9 \Delta^2 \sgn{\bra{\Jflt \cdot \Wflt}}
\sqrt {|\Jflt \cdot \Wflt|/ \flt{\rho}^{1/2}}
 ,
\end{eqnarray}
since \eqref{eq:Wsgs} is unclosed.
$\Wflt = \nabla \times \fav{\VEC{u}}$ is the resolved vorticity.
The closures are motivated by assuming that the modeled cross-helicity 
dissipation rate is a robust proxy
of transfer between kinetic and magnetic energy. 

In addition, we 
include the $\abg$-closure\cite{Yokoi2013} for the 
electromotive force
\begin{eqnarray}
\abg: \qquad
\eta_t = \alpha \flt{\VEC{B}} - \beta \Jflt + \gamma \Wflt
\end{eqnarray}
in our comparison which was recently applied in LES of 
current sheets\cite{2015arXiv151104347W}.
Here, $\beta$ is closed identically to $\ERE$, 
$\gamma = C_{19} t_t \Wsgsdata$ is linked to the SGS cross-helicity and 
$\alpha = C_{20} t_t H$ is connected to the residual helicity 
$H = \flt{\VEC{u}\cdot\VEC{\Omega}} - \fav{\VEC{u}}\cdot \fav{\VEC{\Omega}}
- \bra{\flt{\VEC{B}\cdot\VEC{J}} - \flt{\VEC{B}}\cdot\flt{\VEC{J}}}/\flt{\rho}$.

\paragraph{Scale-similarity} ($\mathrm{SS}$)  closures are characterized by the 
assumption that the tensorial structure at the smallest resolved scales
is similar to the one at the largest unresolved 
scales\cite{1980fpdy.confT....B}.
This motivates the introduction of a second filter 
(a test filter) with a filter width equal to or larger than
the original filter width.
The result of the second filter operation is analogous to
the result of the first filter operation and this allows the 
recovery of  the subgrid-scales.
We use a filter with twice the original filter width, as proposed based on
experimental data\cite{FLM:352690},  and denote
this operation by $\widebreve{\Box}$.
It is understood that mass-weighted 
filtering is applied to all quantities involving $\fav{\VEC{u}}$.
The resulting closures are
\begin{eqnarray}
\SSu: & \qquad &\tu = 
C_{21} \widebreve{\flt{\rho}} \bra{\widebreve{\fav{u_i}\fav{u_j}} 
- \widebreve{\fav{u_i}}\widebreve{\fav{u_j}}} \;, \\
\SSb: & \qquad &\tb = 
C_{22} \bra{\widebreve{\flt{B_i}\flt{B_j}} 
- \widebreve{\flt{B_i}}\widebreve{\flt{B_j}}} \;, \\
\SSemf: & \qquad &\EMF =  
C_{23} \bra{\widebreve{\fav{\VEC{u}}\times\flt{\VEC{B}}} - 
\widebreve{\fav{\VEC{u}}}\times\widebreve{\flt{\VEC{B}}}}
\;.
\end{eqnarray}
It should be noted that these coefficients are introduced in order to 
allow for deviation from model assumptions.
Nevertheless, they are expected to be approximately $1$
due to the self-similarity assumption.
In addition to this, closures for the SGS energies can be extracted 
from these terms directly by means of definition \eqref{eq:EsgsTau}, i.e.
\begin{eqnarray}
\label{eq:EkinsgsSS}
\EkinsgsSS &=&  
\frac{1}{1} C_{21} 
\widebreve{\flt{\rho}} \bra{\widebreve{\fav{u_k}\fav{u_k}} 
- \widebreve{\fav{u_k}}\widebreve{\fav{u_k}}} \;, \\
\EmagsgsSS &=& \frac{1}{2} C_{22}
\bra{\widebreve{\flt{B_i}\flt{B_j}} 
- \widebreve{\flt{B_i}}\widebreve{\flt{B_j}}} \;\text{.} 
\end{eqnarray}

\paragraph{Nonlinear} ($\mathrm{NL}$) closures are structural in nature.
While they are related to other gradient (also tensor-diffusivity) 
closures\cite{Leonard1975237}, they are not based on the expansion of the primary 
quantities, but can be derived
through gradient expansion of the filter kernel\cite{Yeo87}.
In contrast to the other two families, the assumptions here
are rooted in the properties of the filtering
operator and not of turbulence as such.
Truncating the expansion at first order and neglecting the commutator
between mass-weighted filtering and differentiation
lead to the following 
expressions\cite{Vlaykov2016a}:
\begin{eqnarray}
\NLu: & \quad & \tu = 
\frac{1}{12} C_{24} \Delta^2 \flt{\rho} \fav{u}_{i,k} \fav{u}_{j,k} \;, \\
\NLb: & \quad & \tb = 
\frac{1}{12} C_{25} \Delta^2  \flt{B}_{i,k} \flt{B}_{j,k} \;, \\
\NLemfcompr: & \quad & \EMF =  
 \frac{1}{12} C_{26} \Delta^2 \varepsilon_{ijk}  \bigl( \fav{u}_{j,l} \flt{B}_{k,l} \nonumber \\*
 && \qquad 
 - \bra{\ln \flt{\rho}}_{,l} \fav{u}_{j,l} \flt{B}_{k} \bigr) 
\;.
\end{eqnarray}
The electromotive force closure is proposed in our
accompanying paper for the first time.
It goes beyond the previously proposed expression\cite{Balarac2010,Grete2015}
\begin{eqnarray}
\NLemf:  \qquad  \EMF =  
\frac{1}{12} C_{27} \Delta^2 \varepsilon_{ijk} \fav{u}_{j,l} \flt{B}_{k,l}
\end{eqnarray}
by explicitly capturing compressible effects in the second term.
As for the scale-similarity closures, the coefficients 
are external to the closures and meant to capture errors not in-line with 
the closure assumptions.
Thus, values around $1$ are expected.
Again, closures for the SGS energies can readily be written down by
definition~\eqref{eq:EsgsTau} as
\begin{eqnarray}
\EkinsgsNL &=& 
\frac{1}{12} C_{24} \Delta^2 \flt{\rho} \fav{u}_{k,l} \fav{u}_{k,l} \;, \\
\EmagsgsNL &=&  
\frac{1}{12} C_{25} \Delta^2 \flt{B}_{k,l} \flt{B}_{k,l} \;.
\end{eqnarray}
A normalized version of the nonlinear SGS stress tensors
has been proposed in the HD\cite{Woodward2006,Schmidt2011} case and 
in our previous work\cite{Grete2015} for MHD:
\begin{eqnarray}
  \label{eq:tuStarNL}
\NLuE: & \quad & 
  \tu[*] = 2 C_{28} \Ekinsgsdata \left (
  \frac{\fav{u}_{i,k} \fav{u}_{j,k}}{\fav{u}_{l,s} \fav{u}_{l,s}} 
  - \frac{1}{3} \delta_{ij} \right), \\
\NLbE: & \quad &
  \label{eq:tbStarNL}
  \tb[*] = 2 C_{29} \Emagsgsdata \left ( 
  \frac{\flt{B}_{i,k} \flt{B}_{j,k}}{\flt{B}_{l,s} \flt{B}_{l,s	}} 
  - \frac{1}{3} \delta_{ij} \right) .
\end{eqnarray}
Effectively, the strength is locally determined by the SGS energy and 
the structural information is extracted from the unnormalized 
closures $\NLu$ and $\NLb$.
Like the energy-scaled closures within the eddy-dissipation family, 
\eqref{eq:tuStarNL} and \eqref{eq:tbStarNL} are not closed.
For this reason, the $\Ekinsgsdata$ and $\Emagsgsdata$ can be replaced by
the energy closure  \eqref{eq:EnClosures} resulting in
\begin{eqnarray}
  \label{eq:tuStar}
\NLuSStar: & \quad &
  \tu[*] = 2 C_{30} \EkinsgsSStar \left (
  \frac{\fav{u}_{i,k} \fav{u}_{j,k}}{\fav{u}_{l,s} \fav{u}_{l,s}} 
  - \frac{1}{3} \delta_{ij} \right), \\
  \label{eq:tbStar}
\NLbM: & \quad &
  \tb[*] = 2 C_{31} \EmagsgsM \left ( 
  \frac{\flt{B}_{i,k} \flt{B}_{j,k}}{\flt{B}_{l,s} \flt{B}_{l,s	}} 
  - \frac{1}{3} \delta_{ij} \right) .
\end{eqnarray} 

\section{Verification method}
\label{sec:verification}
In a first investigation~\cite{Grete2015} we analyzed the supersonic 
regime in simulations at a resolution of $512^3$ grid points.
Here, we extend the parameter space to include the subsonic and hypersonic 
regime, as well as two additional reference runs at a resolution of $1024^3$ 
grid points.
Furthermore, the functional analysis now goes beyond the turbulent energy cascade --
we also include the cross-helicity cascade and total SGS flux of both resolved
energy and cross-helicity.
Finally, the structural analysis now covers alignment and magnitude of the SGS
vectors, and topological properties of the SGS stresses.

\subsection{Simulations}
\label{ssec:sim}
In total, 15 homogeneous, isotropic turbulence
simulations in a periodic box with varying
sonic Mach number $\Ms$, Alfvenic Mach number $\Ma$ and numerical method were conducted.
All simulations start with uniform initial conditions, i.e. $\rho_0 = 1$, 
$\VEC{u}_0 = \VEC{0}$ 
(these and all following variables are in dimensionless code units) within a
box of length $L = 1$
at resolution of $512^3$ or $1024^3$ grid points.
The initial background magnetic field is uniform in the z-direction and
its magnitude specified by the ratio of thermal to magnetic pressure
$\pb = 2 p/B^2$.
The MHD equations for a compressible fluid are then evolved 
in time using either \Enzo\cite{Enzo2013} or \FLASH\cite{Fryxell2000}.
Statistically stationary turbulence is driven by a 
stochastic forcing field generated by an 
Ornstein-Uhlenbeck process \cite{Pope2000}.
The strength is defined by a characteristic Mach number $V$.
We choose a parabolic forcing profile peaking at wavenumber $k=2$
and a ratio of compressive to solenoidal components 
$\zeta = \abs{\nabla \cdot \VEC{u}}/\|\nabla \VEC{u}\|$ for which we explore
values of $0.5$ and $0.9$.
Details on the forcing can be found in\cite{Schmidt2009,2010A&A...512A..81F} 
and details about individual simulation parameters are listed in 
table~\ref{tab:SimOverview}.
\begin{table*}
	\begin{tabular}{ccccccccc}
		\toprule
		Name & Resolution & Forcing Mach $V$ & Init. $\pb$& 
		$\langle \langle \mathrm{M}_{\mathrm{s}}^2 \rangle^{1/2} \rangle$ & 
		$\langle \langle \mathrm{M}_{\mathrm{a}}^2 \rangle^{1/2} \rangle$ & Code 
		& Riemann solver & $\zeta$ \\
\colrule

 1 & $512^3$	& $0.2$	& $450$	& $0.22$	& $1.95$	& \Enzo 	& HLLD 	& $0.5$\\
 2a & $512^3$	& $0.5$	& $72$	& $0.56$	& $1.85$	& \Enzo 	& HLLD	& $0.5$\\
 2b & $1024^3$	& $0.5$	& $72$	& $0.57$	& $1.81$	& \Enzo 	& HLLD	& $0.5$\\
 3 & $512^3$	& $0.5$	& $8$	& $0.61$	& $1.26$	& \Enzo 	& HLLD	& $0.5$\\
 4 & $512^3$	& $1.0$	& $18$	& $1.17$	& $1.90$	& \Enzo 	& HLLD	& $0.5$\\
 5 & $512^3$	& $1.0$	& $2$	& $1.25$	& $1.27$	& \Enzo 	& HLLD	& $0.5$\\
 6 & $512^3$	& $2.0$	& $5$	& $1.97$	& $2.64$	& \FLASH 	& HLL3R& $0.5$\\
 7a& $512^3$	& $2.0$	& $5$	& $2.46$	& $2.14$	& \Enzo 	& HLL 	& $0.5$\\
 7b& $1024^3$	& $2.0$	& $5$	& $2.55$	& $2.13$	& \Enzo 	& HLL 	& $0.5$\\
 8 & $512^3$	& $2.9$	& $0.25$	& $2.54$	& $0.78$	& \Enzo 	& HLL 	& $0.9$\\
 9 & $512^3$	& $2.9$	& $2.5$	& $2.64$	& $3.11$	& \Enzo 	& HLL 	& $0.9$\\
10 & $512^3$	& $2.9$	& $25$	& $2.68$	& $8.24$	& \Enzo 	& HLL 	& $0.9$\\
11 & $512^3$	& $4.0$	& $1$	& $4.14$	& $2.88$	& \FLASH 	& HLL3R& $0.5$\\
12 & $512^3$	& $10.0$	& $0.2$	& $10.04$	& $2.25$	& \FLASH 	& HLL3R& $0.5$\\
13 & $512^3$	& $20.0$	& $0.05$	& $20.12$	& $2.08$	& \FLASH 	& HLL3R& $0.5$\\
\botrule
	\end{tabular}
	\caption{Overview of analyzed  simulations. The sonic $\Ms$ and 
	Alfvenic $\Ma$ Mach numbers are the temporal means of the spatial RMS numbers 
	over the stationary phase between $2T < t < 5T$ dynamical times.
	In all \Enzo simulations the ideal MHD equations were solved with an
	ideal equation of state.
	For \FLASH a polytropic equation of state and explicit
	viscosity and resistivity (so that $\Re = \Rm = 3780$, see subsection \ref{ssec:sim})
	was used.
	\label{tab:SimOverview}}
\end{table*}
In \Enzo, an open-source fluid
code,
the ideal ($\nu = \eta = 0$) MHD equations are solved with a
MUSCL-Hancock\cite{toro2009riemann} framework, employing second order Runge-Kutta 
integration in time, PLM reconstruction and HLL or HLLD
Riemann solvers\cite{Miyoshi2005315}.
The thermal pressure $p$ is specified by an ideal equation of state
with adiabatic exponent $\kappa = 1.001$ to resemble an isothermal fluid.
In the simulations conducted with the publicly available \FLASH code 
the MHD equations are evolved with explicit\cite{Federrath2011,Federrath2014} 
viscosity $\nu$ and resistivity $\eta$ 
specified via the kinetic Reynolds number 
$\Re = \frac{L_0 V_0}{\nu} = 3780$ and 
the magnetic Reynolds number $\Rm = \frac{L_0 V_0}{\eta} = 3780$.
In all simulations, the characteristic length $L_0 = 0.5L$ is half the box size
due to the forcing profile 
and the characteristic velocity $V_0 = V c_{s,0}$ corresponds to the forcing
Mach number $V$ relative to the initial speed-of-sound $c_{s,0}=1$.
In contrast to \Enzo the gas is kept exactly isothermal by a polytropic 
equation of state.
The chosen numerical scheme consists of second-order integration
in time and space with the HLL3R Riemann solver\cite{Waagan2011}.
For both \Enzo and \FLASH the divergence constraint $\nabla \cdot \VEC{B} = 0$ 
is handled by a divergence cleaning scheme\cite{Dedner2002}.

\begin{figure}[htbp]
\centering
		\includegraphics{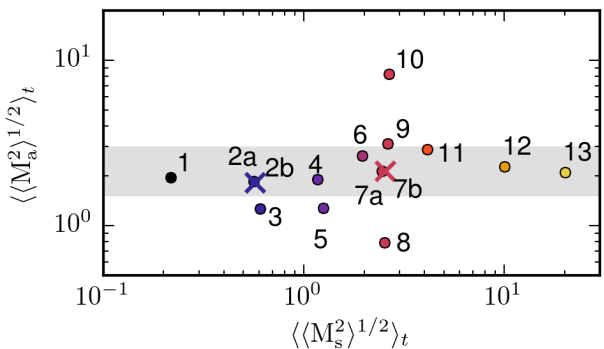}
	\caption{Parameter space covered by the 15 simulations.
		Each marker (circles for a resolution of $512^3$ grid-points
		and  crosses for $1024^3$, respectively)
		corresponds to the respective mean value over
		the stationary phase $2T < t < 5T$ of
		the spatial root mean square Mach numbers.
	Only simulations within the gray area are used in the detailed
	sonic Mach number dependency study.
	Simulation details are given in table \ref{tab:SimOverview}.
	}	
	\label{fig:ParamSpace}
\end{figure}

All simulations initially undergo a transient phase in which the
uniform initial conditions evolve into stationary turbulence.
This phase lasts for $t < 2 T$ dynamical times with $T = 0.5 L / V$.
Afterwards, the gas is evolved for three additional dynamical times
and ten snapshots per dynamical time are captured for the analysis.
The resulting parameter space of the simulations in terms
of the temporal mean ($\langle \Box \rangle_t$) 
sonic $\langle \langle \mathrm{M}_{\mathrm{s}}^2 \rangle^{1/2} \rangle_t$
and Alfvenic $\langle \langle \mathrm{M}_{\mathrm{a}}^2 \rangle^{1/2} \rangle_t$
spatial root mean square ($\langle \Box \rangle$) Mach numbers within
$2T < t < 5T$ is illustrated in figure \ref{fig:ParamSpace}.
Simulations 1, 2a, 4, 6, 7a, 11, 12, 13 within the gray area have
$\langle \langle \mathrm{M}_{\mathrm{a}}^2 \rangle^{1/2} \rangle_t \approx 3$
and are therefore used for a $\Ms$-dependency analysis of the different closures.

\subsection{Reference quantities}
\label{ss:refQuant}
\begin{figure}[htbp]
\centering
		\includegraphics{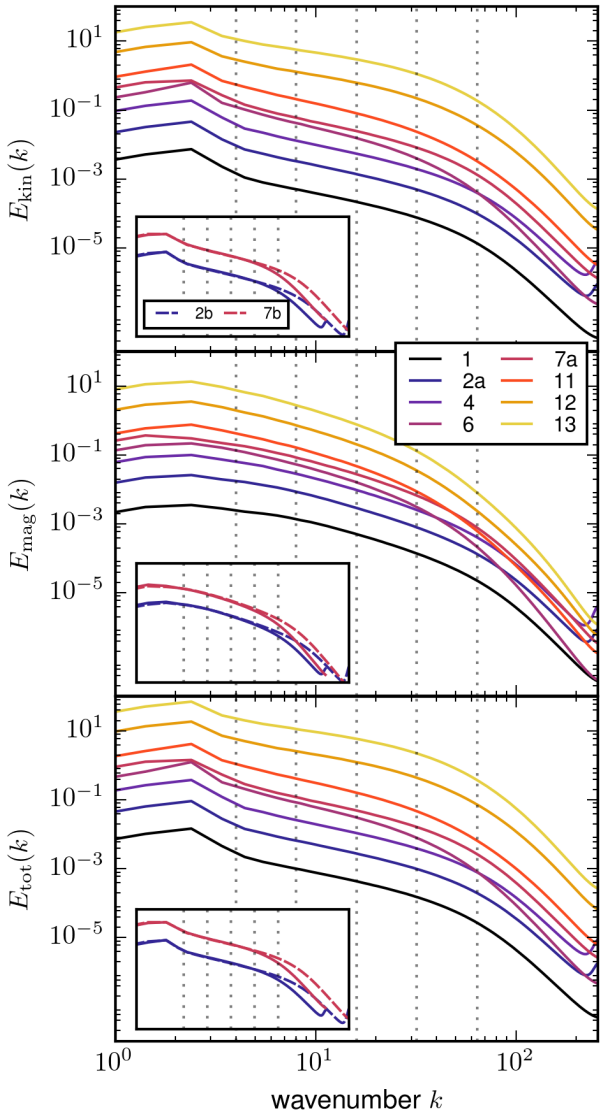}
	\caption{Mean ($2T < t < 5T$) power spectra of the simulations. 
		Kinetic energy is based on the Fourier transform of $\sqrt{\rho}\VEC{u}$.
		The dashed vertical lines indicate the filter widths ($k=4,8,16,32,64$)
		we are using during the analysis. The insets highlight the
		extended power-law regime of the $1024^3$ runs (2b and 7b, dashed lines) over the
		corresponding $512^3$ runs (2a and 7a, solid lines). Simulation details are listed
		in table \ref{tab:SimOverview}.
		\label{fig:PowerSpecAllRuns}}
\end{figure}

In order to assess the quality and performance of the different closures 
we conduct functional and structural \aprio tests.
In \aprio testing a test filter is applied to high resolution data 
to mimic the effect of limited resolution. 
The scales below the test filter are treated as unresolved scales.
Owing to the explicit filtering we not only obtain filtered quantities 
intended to resemble the resolved scales, but also
retain the sub-filter quantities intended to resemble the unresolved scales.
This allows the exact calculation of SGS quantities.
In the context of LES three different filter kernels are typically 
used\cite{Sagaut2006}: the box, the Gaussian, and the sharp spectral
filter.
For the majority of our analysis we use a Gaussian filter
with a characteristic filter scale at a wavenumber $k = 16$ 
for several reasons.
Firstly, $k=16$ is within a power-law regime of the energy 
spectra (cf. figure~\ref{fig:PowerSpecAllRuns}),
which satisfies the assumption of the eddy-viscosity
and scale-similarity type closures.
Secondly, it is sufficiently far away from the forcing scale 
$k=2$ where the dynamics of the forcing are expected to be dominant.
Thirdly, it also does not fall above the high-$k$ drop-off in the
spectrum, caused by viscous and numeric dissipation,
which contaminates turbulent dynamics \cite{Kitsionas2009}.
The mean spectra within the stationary regime ($2T < t <5T$) 
of the simulations are illustrated
in figure~\ref{fig:PowerSpecAllRuns}, where we also highlight 
the filter positions.
In addition to filtering at $k=16$ we also probe the closures with filter 
scales at $k=4,8,32,64$ to investigate the dependence of the result on the
chosen scale.
Moreover, we verify the results based on Gaussian filtering
against a box filter.
Given that we analyze compressible data, we do not employ a 
sharp spectral filter, which can produce negative resolved densities, and 
SGS stresses that violate realizability\cite{Vreman1994}.

The first category of tests, functional tests, probe the ability of 
closures to reproduce a particular (physical) property.
In addition to this, functional tests can eliminate co-ordiate frame dependence 
by reduction to scalar diagnostics, e.g.
of six SGS stress tensor or three EMF vector components.
Historically, the most frequently used reference quantity is the 
turbulent energy flux, i.e. the cascade term
\begin{eqnarray}
\SE = \tau_{ij} \fav{\mathcal{S}}_{ij} + \EMFdata \cdot \flt{\VEC{J}} \;.
\end{eqnarray}
It encodes the local exchange between resolved and unresolved energy
and is connected to the turbulent energy cascade.
However, as it was recently shown\cite{Vlaykov15}, the total energy flux term 
\begin{eqnarray}
\label{eq:FE}
\FE = - \fav{\VEC{u}} \cdot \bra{\nabla \cdot \tau} + \flt{\VEC{B}} \cdot \nabla \times \EMFdata 
\end{eqnarray}
is more strongly influenced by the transport terms 
${\nabla \cdot \bra{\fav{\VEC{u}} \cdot \tau + \flt{\VEC{B}}\times \EMFdata}}$
rather than 
the cascade term $\SE$ in our simulations.
Furthermore, in MHD there are additional conserved quantities such as
cross-helicity, ${W = \VEC{u} \cdot \VEC{B}}$, which are, in the context of LES,
also governed by resolved and subgrid-scale evolution equations\cite{Vlaykov15}.
The exchange of cross-helicity across the filter scale is analogous to
the energy one, with cross-helicity flux 
\begin{eqnarray}
\SW = \tau_{ij} \bra{\flt{B_{i}}/\flt{\rho}}_{,j}
+ \EMFdata \cdot \fav{\VEC{\Omega}}
\;.
\end{eqnarray}
Again, the total cross-helicity term
\begin{eqnarray}
	\FW = - \flt{\VEC{B}}/\flt{\rho} \cdot \bra{\nabla \cdot \tau} + \fav{\VEC{u}} \cdot \nabla \times
\EMFdata
\;,
\end{eqnarray}
is dominated by the transport and not the cascade contribution\cite{Vlaykov15}.
In the following we are going to analyze all four (pseudo-)scalars as 
each of them may play a crucial role in different dynamical regimes, and
systematic differences between results from total and cascade fluxes may
indicate the importance of the differentiation commutator\cite{Vlaykov2016a}.
Specifically, we conduct nonlinear least-square 
minimization\cite{newville_2014_11813} between 
data and closure.
This automatically produces the best coefficient $C_{\Box}$
for each snapshot and closure individually. 
Eventually, we calculate the Pearson correlation coefficient as an
overall measure of accuracy.
While these correlations probe the spatially local performance of the closures, 
we also analyze a global indicator.
In particular, we look at the average SGS dissipation, i.e. the total 
$\SE$ for each snapshot, and examine the contributions of the individual
components.

The performed structural tests 
start with a topological analysis.
We use the geometric invariants of second-rank tensors to compare the topology
of the deviatoric SGS Reynolds $\tudata[*]$ and Maxwell $\tbdata[*]$
stress tensors for data and closure.
The characteristic polynomial of a second-rank
tensor $\mathcal{T}$ is\cite{Dallas2013}
$\lambda_i^3 + P \lambda_i^2 + Q \lambda_i + R = 0$ with eigenvalues 
$\lambda_i$ and invariants
\begin{eqnarray}
P &=& -\tr\bra{\mathcal{T}} = - \bra{\lambda_1 + \lambda_2 + \lambda_3}\;,\\
Q &=& \frac{1}{2} \bra{P^2 - \tr{\bra{\mathcal{T}^2}}} 
= \lambda_1\lambda_2 + \lambda_2\lambda_3 + \lambda_3\lambda_1 \;,\\
R &=& -\det\bra{\mathcal{T}} = -\lambda_1\lambda_2\lambda_3 \;.
\end{eqnarray}
Both tensors, $\tudata[*]$ and $\tbdata[*]$, are traceless, so $P=0$.
Furthermore, they are symmetric.
Thus, $Q$ is negative definite and the three eigenvalues 
$\lambda_1 \geq \lambda_2 \geq \lambda_3$ are real.
Therefore, only two eigenvalue combinations are possible.
On the one hand, sheet-like structures with $R>0$ are 
produced by expansion in two dimensions 
($\lambda_1,\lambda_2 > 0$),
and contraction in the third dimension ($\lambda_3 < 0$).
On the other hand,  tube-like structures with $R<0$ are produced by
expansion in one dimension ($\lambda_1 > 0$), and contraction in two dimensions 
($\lambda_2,\lambda_3 <0$).

Given that all closures enter the primary equations ultimately in vectorial
form, we also asses their geometrical performance.
For this reason, we compare the alignment of the data vector, 
e.g. $\nabla \cdot \tudata[*]$,
with the corresponding closure vector, i.e. $\nabla \cdot \tu[*]$.
Moreover, we compare their respective magnitudes.
Ideally, the modeled SGS vector will point in the identical direction
as the data vector
($\cos \left ( \nabla \cdot \tu[*], \nabla \cdot \tudata[*] \right ) = 1$),
and will be with identical magnitude 
($|\nabla \cdot \tu[*]|/|\nabla \cdot \tudata[*]| = 1$).

\section{Results}
\label{sec:results}

\subsection{Functional analysis: overview and $\Ms$ dependency}
\begin{figure}[htbp]
\centering
\includegraphics{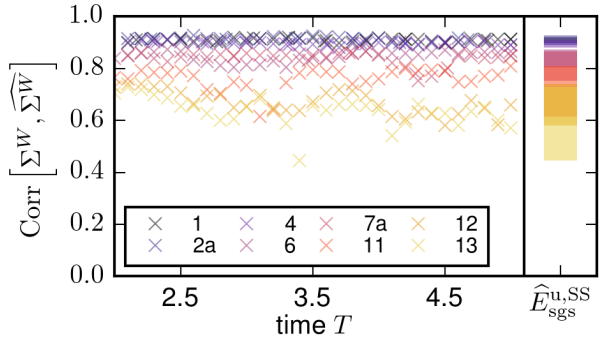}
\caption{Illustration of using correlations from individual simulation snapshots (left)
to create a bar plot (right). 
Each marker on the left side corresponds to the correlation coefficient from one snapshot
of the color-coded simulations from the subsonic (dark) to the hypersonic (bright) regime.
The correlation coefficient is always calculated for only one reference quantity (here, 
the cross-helicity cascade flux $\SW$) with one closure (here, the kinetic SGS energy 
of the scale-similarity family $\EkinsgsSS$).
Each colored bar on the right side spans the range of variation from the minimum to the
maximum correlation value over all snapshots of one simulation. 
}
\label{fig:CorExplanation}
\end{figure}
We start our functional analysis by evaluating the performance of the
different closures for the isotropic parts of the SGS stresses, i.e.
by definition \eqref{eq:EsgsTau}, the SGS energies 
$\tau_{ij}^{\Box} = 2/3 \delta_{ij} \Esgsdata^{\Box}$.
Figure~\ref{fig:CorExplanation} illustrates the creation of the Mach number
dependent bar plots we use in this section for one sample quantity.
We use the kinetic SGS energy closure of the scale-similarity family
$\EkinsgsSS$ \eqref{eq:EkinsgsSS}, and compute the correlation
$ \mathrm{Corr} \left[ \SW , \widehat{\SW} \right ]$ 
of the contribution to cross-helicity cascade term $\SW$ based on closure and
exact SGS energy expression $\Ekinsgsdata$ \eqref{eq:SGSen}, i.e.
\begin{equation}
\mathrm{Corr} \left[ 
\frac{2}{3}  \delta_{ij} \Ekinsgsdata \bra{\frac{\flt{B_i}}{\flt{\rho}}}_{,j},
\frac{2}{3}  \delta_{ij} \EkinsgsSS \bra{\frac{\flt{B_i}}{\flt{\rho}}}_{,j}
\right] \;.
\end{equation}
This is done for each snapshot of all simulations.
Then, we take the
minimum and maximum value of each simulation separately
to determine the vertical extent of the color-coded bars 
in the right panel of the figure.
In this example it is clear that the cross-helicity cascade is well modeled in
the subsonic regime (with correlations above 0.8) and tends to perform worse 
in the hypersonic regime (going down to almost 0.4).

\begin{figure*}[htbp]
\centering
\subfigure[Correlations of different SGS energy closures, i.e. the isotropic component of the SGS stress tensors. ]{
\includegraphics{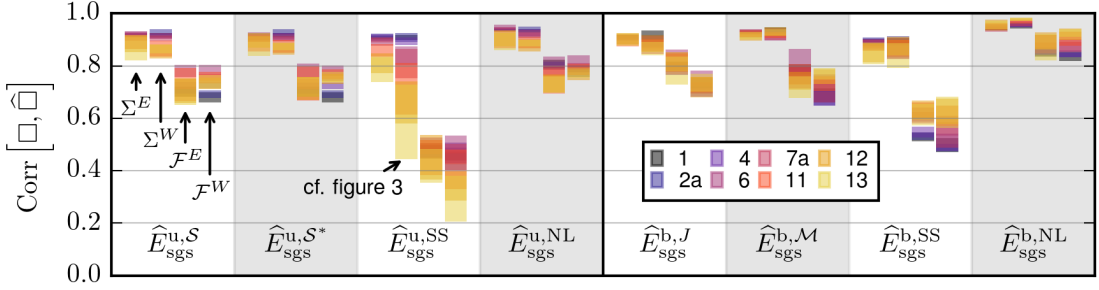}
\label{fig:EnergyOverview}
}

\subfigure[Correlations of different traceless SGS Reynolds stress $\tau^{\mathrm{u}*}$ closures. ]{
\includegraphics{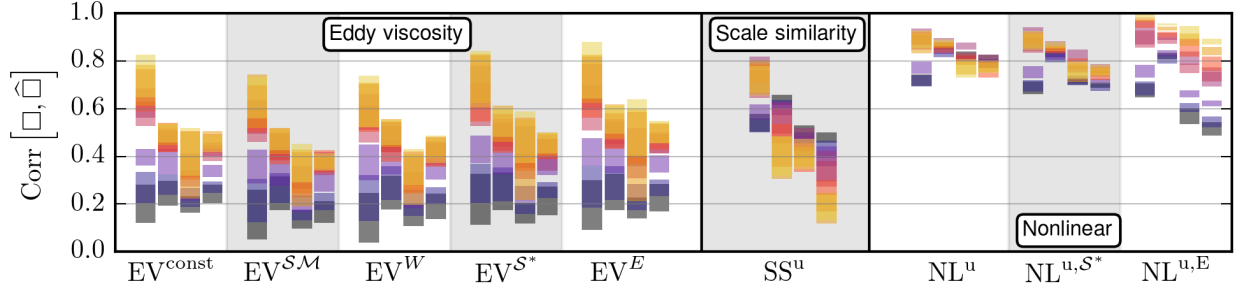}
\label{fig:tauUStarOverview}
}

\subfigure[Correlations of different traceless SGS Maxwell stress $\tau^{\mathrm{b}*}$ closures.]{
\includegraphics{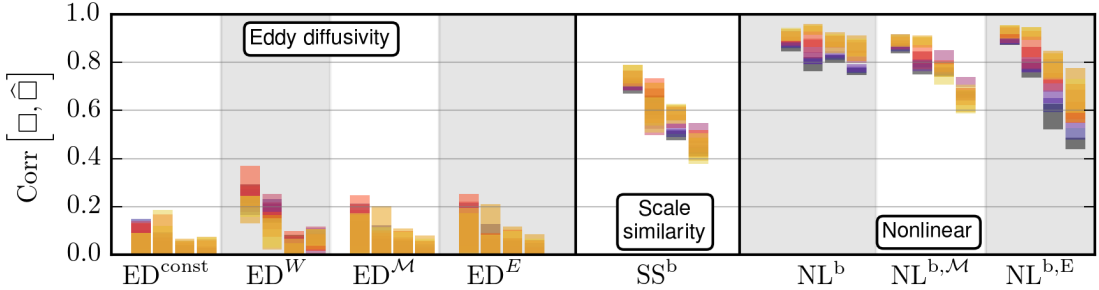}
\label{fig:tauBStarOverview}
}

\subfigure[Correlations of different electromotive force $\mathcal{E}$ closures.]{
\includegraphics{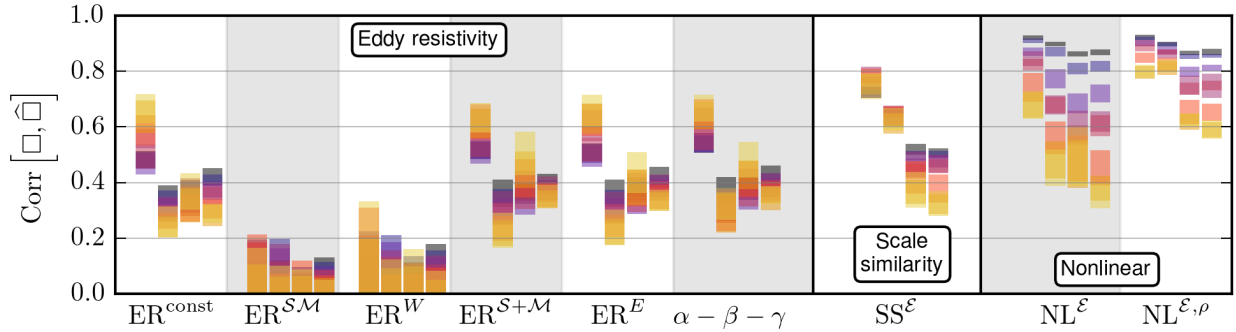}
\label{fig:EMFOverview}
}

\caption{
Correlations between closure and data for all reference fluxes.
For each closure the four colored bars 
	(from left to right:
	energy  $\Sigma^E$ and cross-helicity $\Sigma^W$ cascade, and
	total energy $\mathcal{F}^E$ and cross-helicity $\mathcal{F}^W$ flux)
illustrate the maximum range of correlation 
split by simulation. 
A detailed explanation of the colored bars is given in figure~\ref{fig:CorExplanation}.
Subsonic runs are towards the dark end and supersonic
at the bright end of the palette (cf. figure \ref{fig:PowerSpecAllRuns}).
All simulations have been filtered at $k=16$. 
The x-axis labels denote the different closure identifiers as introduced in section 
\ref{sec:closures}.
\label{fig:ModelOverview}
}
\end{figure*}

The results for all energy closures and all reference quantities are shown
in figure \ref{fig:EnergyOverview}.
In general, all closures perform very well with respect to the cascade fluxes.
A notable exception is the already mentioned cross-helicity cascade correlation
of the kinetic scale-similarity model $\EkinsgsSS$, which has a strong
$\Ms$ dependency.
In addition to this, it can be seen that the total flux terms are generally
less well represented than the cascade terms.
Furthermore, there is practically no difference between modeling the 
eddy-viscosity/diffusivity energies based on realizability conditions
($\EkinsgsSStar$ and $\EmagsgsM$) and the equilibrium approach
($\EkinsgsS$ and $\EmagsgsJ$). 
Overall, with a slight advantage over the eddy-viscosity closures, 
the nonlinear closures perform best with generally high correlations 
(\textgreater$0.7$ across the entire parameter space) and very limited $\Ms$
dependency.
The median across all simulations of the free coefficient value of each closure
is listed in table \ref{tab:valueoverview} including bounds given by the
interquartile range (IQR).
For reference we also provide more detailed data tables as supplemental material\cite{suppmat}.
All SGS energy closures exhibit only a very limited spread over the tested parameter space
with IQRs within a factor of 2 around the median.
These results also hold (not shown) for direct fits, i.e. 
$\mathrm{Corr} \left[E_{\mathrm{sgs}}^\Box,\widehat{E}_{\mathrm{sgs}}^\Box\right]$,
of the kinetic $\Ekinsgsdata$, magnetic $\Emagsgsdata$, and total
$\Esgsdata$ energies.
\begin{table}
\caption{\label{tab:valueoverview} Median correlation and coefficient values
over all $512^3$ simulations filtered at $k=16$ with
lower and upper bound given by interquartile range of all data.
Detailed data tables including results split by reference quantity and 
min-/maximum values can be found in the supplementary material\cite{suppmat}.}
\begin{tabular}{@{}lll}
	\toprule 
	ID	& 
	$\mathrm{Corr} \left [ \Box, \widehat{\Box} \right ]$ &
	Coefficient\\
	\colrule
$\EkinsgsS$& $0.83^{+0.082}_{-0.094}$ & $C_{4}=0.036^{+0.014}_{-0.0074}$  \\
$\EkinsgsSStar$& $0.84^{+0.068}_{-0.098}$ & $C_{2}=0.038^{+0.022}_{-0.005}$  \\
$\EkinsgsSS$& $0.59^{+0.3}_{-0.11}$ & $C_{21}=1^{+0.19}_{-0.35}$  \\
$\EkinsgsNL$& $0.85^{+0.077}_{-0.058}$ & $C_{24}=1.2^{+0.53}_{-0.19}$  \\
\colrule
$\EmagsgsJ$& $0.83^{+0.057}_{-0.084}$ & $C_{3}=0.043^{+0.021}_{-0.0065}$  \\
$\EmagsgsM$& $0.87^{+0.054}_{-0.13}$ & $C_{1}=0.045^{+0.028}_{-0.0058}$  \\
$\EmagsgsSS$& $0.79^{+0.093}_{-0.23}$ & $C_{22}=1.1^{+0.26}_{-0.36}$  \\
$\EmagsgsNL$& $0.93^{+0.018}_{-0.073}$ & $C_{25}=1.3^{+0.43}_{-0.1}$  \\
\colrule
$\EVconst$& $0.4^{+0.079}_{-0.12}$ & $C_{5}=0.096^{+0.13}_{-0.074}$  \\
$\EVSM$& $0.35^{+0.081}_{-0.1}$ & $C_{8}=0.011^{+0.0061}_{-0.0057}$  \\
$\EVW$& $0.39^{+0.092}_{-0.12}$ & $C_{10}=0.024^{+0.008}_{-0.008}$  \\
$\EVSStar$& $0.43^{+0.091}_{-0.13}$ & $C_{16}=0.0085^{+0.0031}_{-0.0031}$  \\
$\EVE$& $0.44^{+0.089}_{-0.15}$ & $C_{13}=0.041^{+0.017}_{-0.021}$  \\
\colrule
$\EDconst$& $0.02^{+0.016}_{-0.012}$ & $C_{6}=0.00071^{+-0.006}_{-0.006}$  \\
$\EDW$& $0.089^{+0.1}_{-0.042}$ & $C_{11}=-0.0066^{+0.0025}_{-0.0067}$  \\
$\EDM$& $0.026^{+0.021}_{-0.012}$ & $C_{17}=0.00014^{+0.00014}_{-0.00038}$  \\
$\EDE$& $0.027^{+0.02}_{-0.014}$ & $C_{14}=0.00055^{+0.00093}_{-0.002}$  \\
\colrule
$\ERconst$& $0.35^{+0.092}_{-0.053}$ & $C_{7}=0.14^{+0.054}_{-0.11}$  \\
$\ERSM$& $0.032^{+0.024}_{-0.017}$ & $C_{9}=-0.00055^{+0.0011}_{-0.0015}$  \\
$\ERW$& $0.042^{+0.035}_{-0.024}$ & $C_{12}=-0.0014^{+0.0021}_{-0.0039}$  \\
$\ERSpM$& $0.36^{+0.11}_{-0.057}$ & $C_{18}=0.0096^{+0.0068}_{-0.0035}$  \\
$\ERE$& $0.36^{+0.1}_{-0.056}$ & $C_{15}=0.035^{+0.025}_{-0.013}$  \\
$\abg$& $0.37^{+0.11}_{-0.049}$ & $\left\{ \begin{tabular}{@{\ }l@{}} $C_{20}=-0.0026^{+0.0018}_{-0.0043}$ \\
$C_{15}=0.033^{+0.028}_{-0.0087}$  \\
$C_{19}=-0.00017^{+0.0058}_{-0.0079}$  \\
\end{tabular}\right.$  \\
\colrule
$\SSu$& $0.49^{+0.11}_{-0.072}$ & $C_{21}=0.67^{+0.16}_{-0.23}$  \\
\colrule
$\SSb$& $0.58^{+0.081}_{-0.084}$ & $C_{22}=0.9^{+0.25}_{-0.43}$  \\
\colrule
$\SSemf$& $0.55^{+0.13}_{-0.084}$ & $C_{23}=0.89^{+0.098}_{-0.18}$  \\
\colrule
$\NLu$& $0.82^{+0.038}_{-0.029}$ & $C_{24}=0.98^{+0.081}_{-0.19}$  \\
$\NLuSStar$& $0.77^{+0.069}_{-0.038}$ & $C_{30}=0.032^{+0.0026}_{-0.0052}$  \\
$\NLuE$& $0.81^{+0.078}_{-0.13}$ & $C_{28}=0.52^{+0.09}_{-0.12}$  \\
\colrule
$\NLb$& $0.85^{+0.029}_{-0.038}$ & $C_{25}=1.1^{+0.19}_{-0.063}$  \\
$\NLbM$& $0.77^{+0.065}_{-0.074}$ & $C_{31}=0.039^{+0.0093}_{-0.0052}$  \\
$\NLbE$& $0.76^{+0.11}_{-0.14}$ & $C_{29}=0.52^{+0.21}_{-0.21}$  \\
\colrule
$\NLemf$& $0.7^{+0.13}_{-0.13}$ & $C_{27}=1.2^{+0.14}_{-0.11}$  \\
$\NLemfcompr$& $0.84^{+0.04}_{-0.072}$ & $C_{26}=1^{+0.11}_{-0.3}$  \\
\colrule
\end{tabular}
\end{table}

The correlations of all four functional reference quantities for
the traceless SGS Reynolds stress $\tu[*]$ are 
depicted in figure~\ref{fig:tauUStarOverview}.
All eddy-viscosity type closures $\mathrm{EV}^{\Box}$ are very similar and 
insensitive of the scaling chosen.
Even though the correlations for all snapshots within a single
simulation do not vary much, there is a substantial difference between
the simulations.
Correlations are typically below $0.2$ in the subsonic regime
whereas they can reach \textgreater$0.8$ in the highly supersonic regime.
This ordering is present in all reference quantities for the $\mathrm{EV}$ closures.
The scale-similarity $\SSu$ closure also exhibits this behavior for the turbulent
energy cascade $\SE$ even though the lower bound in the subsonic regime is 
much better, $\approx 0.5$.
However, the correlations of the other fluxes, $\SW$, $\FE$ and $\FW$, have 
the opposite ordering.
The most extreme case of $\FW$ 
spreads from $\approx 0.5$ in the subsonic regime
to correlations as low as $0.1$ for the $\Ms \approx 20$ simulation. 
The nonlinear family $\mathrm{NL}$ is closest to the data in general.
Again, we observe an ordering with $\Ms$ but the spread is much more
constrained and for $\NLu$ the correlations are consistently above $0.7$.
Here, the scaling only further separates individual simulations with
supersonic simulations slightly improving and subsonic simulations
becoming worse on average.

Generally, the results for the traceless SGS Maxwell stress $\tb[*]$,
as shown in figure~\ref{fig:tauBStarOverview}, are very similar to 
those for $\tu[*]$.
Again, the nonlinear family has the best performance and different
normalizations for $\mathrm{NL}$ cause a wider spread.
The scale-similarity closure $\SSb$ is slightly worse with
best correlations up to $0.8$ for $\SE$ and worst -- $0.4$ for
$\FW$.
Most striking is the poor performance of all eddy-diffusivity 
($\mathrm{ED}$) closures.
Independent of normalization and simulation the correlations
barely reach $0.4$ with the majority of snapshots (93\%) being
below $0.2$ for all reference quantities.

Finally, the findings for the electromotive force $\EMFdata$ 
are much more diverse.
Firstly, within the eddy-resistivity ($\mathrm{ER}$) family,
scaling by cross-helicity leads to poor correlations
(99\% snapshots below $0.2$).
However, $\ERconst$ and energy scalings
($\ERSM$ and $\ERE$) provide reasonable correlations 
(from $0.5$ for low $\Ms$ to $0.7$ for high $\Ms$)
for the turbulent energy cascade $\SE$, but are less
effective (\textless$0.5$) for $\SW$, $\FE$ and $\FW$.
In addition, there is practically no difference between these
scalings and the addition of the two extra terms in the
$\abg$ closure.
Secondly, the scale-similarity closure $\SSemf$ performs
similar to the reasonable $\mathrm{ER}$ closures  
with respect to the total terms $\FE$ and $\FW$.
However, it performs much better for the cascade terms
with correlations for $\SW$ of $\approx 0.65$ and for
$\SE$ of $\approx 0.75$ without significant $\Ms$ dependence.
Thirdly, the effect of the compressible extension of the nonlinear closure
$\NLemfcompr$ becomes apparent when comparing the results for different
simulations.
While there is practically no difference between $\NLemf$ and $\NLemfcompr$
in the subsonic regime (correlations \textgreater$0.9$ for all quantities),
the shortcomings of $\NLemf$ in the highly
supersonic regime are apparent.
Correlations of $\approx 0.4$ for $\SW$, $\FE$ and $\FW$ in the 
$\Ms \approx 20$ simulation can be improved by the additional term
in $\NLemfcompr$ to $\approx 0.6$ for $\FE$ and $\FW$, and even up to 
$\approx 0.8$ for $\SW$.
The improvements for $\NLemfcompr$ are more pronounced in the cascade terms 
(with a spread of 0.8-0.9) than in the total flux terms (with a spread of 0.6-0.9).
Here, the additional differentiation commutator\cite{Vlaykov2016a} might further
increase the correlations in the high-$\Ms$ regime.
The overall trend that the nonlinear closures are
better correlated with the data than the scale-similarity or
eddy-resistivity closures continues for the electromotive force as well.

Furthermore, as listed in table~\ref{tab:valueoverview}
closures that exhibit a generally high correlation show the least spread
in their free coefficient values $C_{\Box}$ and vice versa.
For example, $\NLu$, with a median correlation of $0.82$, has a spread
in the coefficient value of \textless$20\%$.
In contrast to this, $\EDE$, with a median correlation of $0.027$, has median
coefficient of effectively $0$ because it takes both negative and positive values.
It should be noted that all scale-similarity closures and the unnormalized 
nonlinear closures have coefficients of $C_{\Box} \approx 1$, as expected analytically.
Finally, the common coefficient $C_{15}$, which the 
$\abg$ and $\ERE$ closures share, is essentially identical, 
while the two additional terms in the $\abg$ closure are effectively canceled 
by their free coefficients $C_{19}, C_{20} \approx 0$.
This also explains their identical behavior in correlations.

\subsection{Functional analysis: filter widths and kernel shapes}
\begin{figure*}[htbp]
\centering
		\subfigure[\label{fig:resFilterStudytauU}
	SGS Reynolds stress closures {\protect $\tu[*]$}]{
			\includegraphics{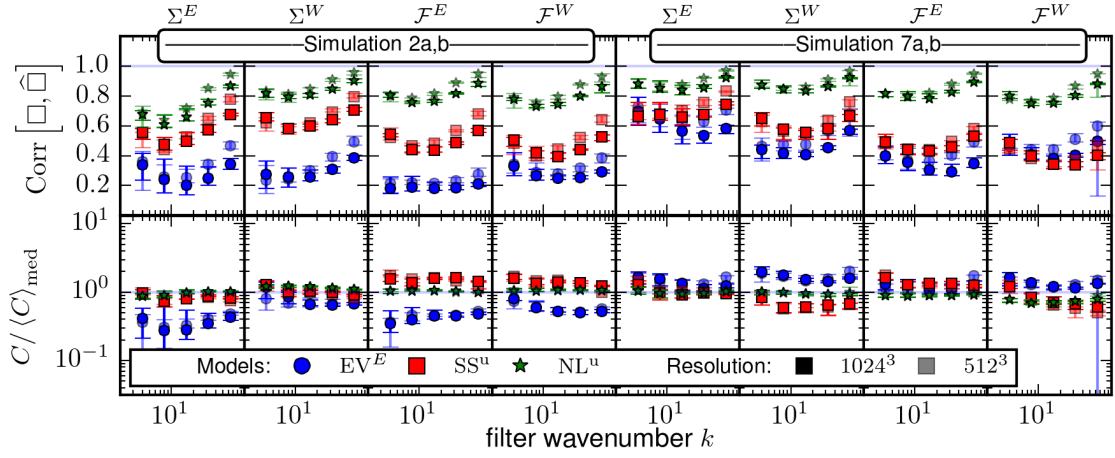}
		}
		\subfigure[\label{fig:resFilterStudytauB}
	SGS Maxwell stress closures {\protect $\tb[*]$}]{
			\includegraphics{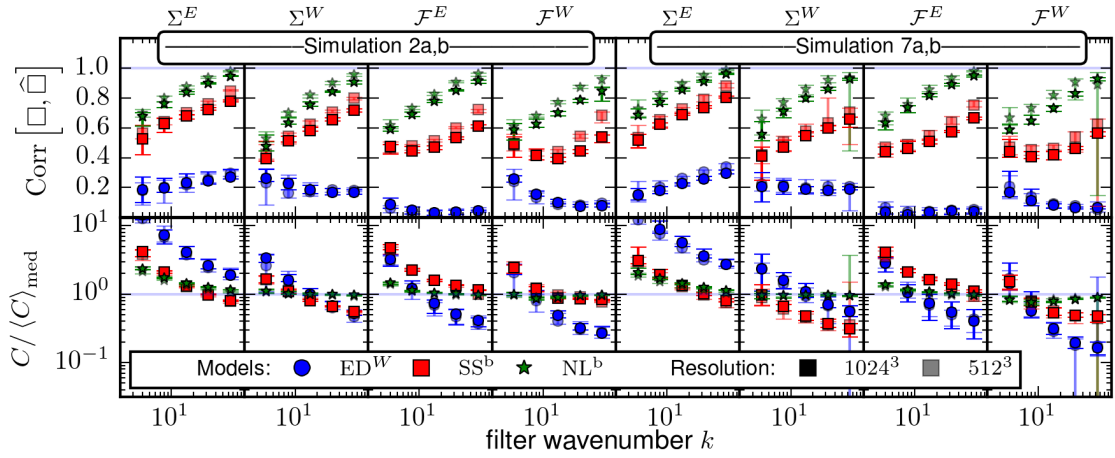}
		}
			\subfigure[\label{fig:resFilterStudyEMF}
	Electromotive force closures $\EMF$]{
			\includegraphics{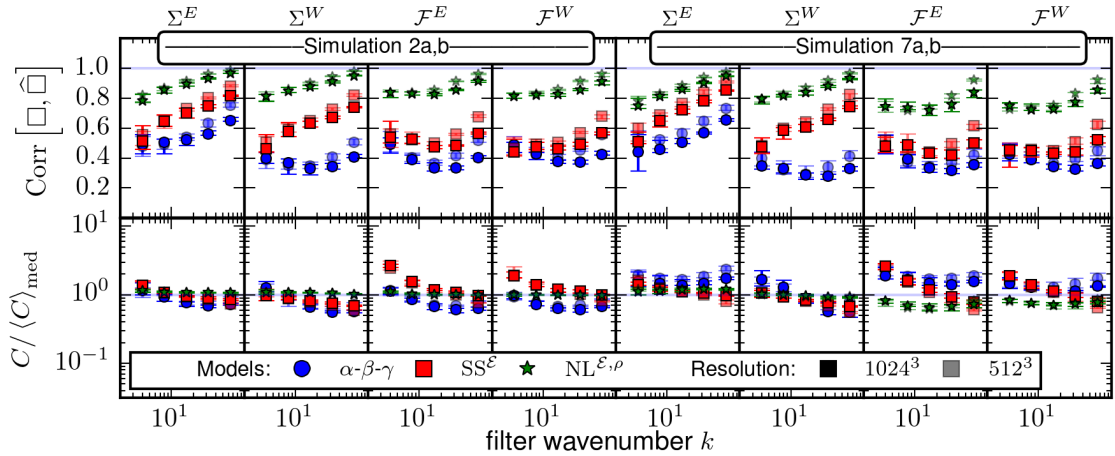}
		}
	\caption{
Comparison of the median correlation (top row in each plot) and coefficient 
(bottom row) value at different filter wavenumbers $k=4,8,16,32,64$ and simulation 
resolutions $512^3$ (transparent) and $1024^3$ (opaque) for subsonic simulation 2a,b
and supersonic simulation 7a,b.
The error bars illustrate the respective minimum and maximum values.
Each column corresponds to results of fitting one reference quantity $\SE$, $\SW$, 
$\FE$ or $\FW$, and each marker represents the median value over snapshots
at $t=\{2,2.5,3,3.5,4,4.5,5\}T$ of the particular simulation.
The coefficient values are normalized to the respective median value over the
snapshots of both simulations and at all filter widths at a given resolution.
}
\label{fig:resFilterStudy}
\end{figure*}
In the last section we saw that the differences in correlations for functional tests
are most pronounced between closure families and that normalization within a 
family itself is subdominant.
For this reason, we continue our analysis with the best performing closure
of each family.
In this section we verify that the results shown in the last section from 
simulations at a resolution of $512^3$ filtered at $k=16$ 
do not substantially change with resolution
and we investigate how the different closures react to 
the chosen filter scale.

Figure \ref{fig:resFilterStudy} illustrates the comparison of correlation and 
coefficient values among four simulations (2a,b and 7a,b)
that differ in driving
(subsonic and supersonic) and resolution ($512^3$ and $1024^3$).
Furthermore, we apply the filter at different scales $k=4,8,16,32,64$.
The extreme cases, $k=4$ and $k=64$, are very close to the forcing regime or  
already in the dissipation regime\cite{Kitsionas2009}, respectively.
Generally, we confirm the observed ordering in correlations among closure
families described in the last section.
Independent of resolution and filter width, the nonlinear closures outperform
the scale-similarity and eddy-viscosity type closures.
On average the difference in both correlations and coefficient values
between the $512^3$ and $1024^3$ simulations are below 7\% at $k=16$. 
Furthermore, all closures typically achieve higher correlations 
($\approx25\%$ while filtering at $k=64$ compared to $k=16$) 
towards the high-$k$ end and the correlations from $512^3$ simulations at $k>16$ 
tend to be higher than from simulations at $1024^3$.
This is not surprising.
On the one hand, the amount of subgrid-scale dynamics that needs to be modeled
is reduced with increasing filter wavenumber.
One the other hand, there is less physical information at high $k$ for lower
resolutions.
Nevertheless, for some cases there are more subtle differences with respect to filter scale,
which we describe in the following.

In figure \ref{fig:resFilterStudytauU} the best closures within each family
for the SGS Reynolds stress are shown, i.e. $\EVE$, $\SSu$ and $\NLu$.
The overall correlation, depending on filter scale $k$,
for each model and reference quantity has a very shallow U-shape.
Compared to $k=16$, the correlations are $\approx 6\%$ higher at $k=4$ and
$\approx 30\%$ higher at $k=64$, respectively.
The slight increase at $k=4$ might be attributed to the proximity to the 
forcing scale $k\approx2$, which is completely resolved.
Thus, the largest unresolved scales of $\tudata$
might see an imprint of the (resolved) forcing and lack SGS turbulent dynamics, 
which, in turn, renders specific SGS modeling unnecessary and increases the correlation. 
The observed systematic differences in correlations with varying $k$ are 
generally not present in the coefficient values.
However, the values vary to different extents within each family and reference
quantity.
While the mean deviation from the median coefficient over all reference quantities,
filter widths and snapshots is only 10\% for the nonlinear closure $\NLu$,
it varies by $47\%$ for the eddy-viscosity reference closure
$\EVE$.
Compared to the results of $\tbdata[*]$ in the next paragraph, this is still
acceptable, even though we find systematically lower coefficient 
at $\Ms \approx 0.6$ compared to $\Ms \approx 2.5$.

The SGS Maxwell stress results depicted in figure~\ref{fig:resFilterStudytauB} 
show a strong filter scale dependency of the closure coefficient
for the scale-similarity $\SSb$ and eddy-diffusivity $\EDW$ closure.
The coefficients are larger for small $k$ and decrease with increasing $k$
spanning almost two orders-of-magnitude.
Only the nonlinear closure $\NLb$ keeps a rather constant value with
deviations of $17\%$ on average.
The correlations, on the other hand, show a systematic increase 
with $k$ for $\NLb$ in all reference quantities.
This might be ascribed to the absence of a direct forcing term acting 
on the magnetic field.
Similar behavior is also present in $\SSb$ with the slight difference of
a plateau for $k\leq16$ in the total flux quantities $\FE$ and $\FW$.
Finally, the eddy-diffusivity closure never reaches a correlation higher than
$0.36$ over the entire parameter space.

The different closures for the electromotive force $\EMFdata$
are closer to each other as illustrated in 
figure~\ref{fig:resFilterStudyEMF}.
Here, both $\NLemfcompr$ and $\SSemf$ exhibit strictly increasing
correlation values with $k$ for the cascade terms $\SE$ and $\SW$ and
a plateau for $k\leq16$ in the total flux terms $\FE$ and $\FW$.
The coefficient values for all $\EMFdata$ closure are less widely spread.
The $\abg$ closure has a variation of $37\%$ around the median over all data 
whereby we only take the dominant $\beta$ term into account.
The $\SSemf$ closure has a variation of $47\%$ and the nonlinear closure is
effectively constant with a spread of only $16\%$.

\begin{figure}[htbp]
	\begin{center}
		\includegraphics{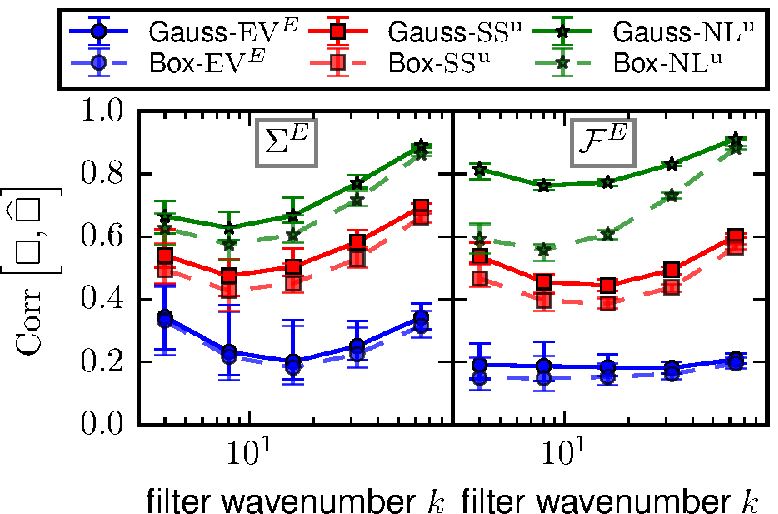}
	\end{center}
	\caption{Correlations of the energy cascade, $\SE$, and total
		energy, $\FE$, flux of different deviatoric kinetic SGS stress
		closures for different filter widths 
		and kernels (box~-~- and Gaussian~---) in subsonic 
		simulation 2b.
		Markers indicate the median and the error bars show the minimum
		and maximum value over time.
	}
	\label{fig:filterAnalysis}
\end{figure}
Finally, the differences between using a Gaussian and a box kernel for the 
analysis are illustrated in figure~\ref{fig:filterAnalysis}.
Two trends can be observed for the kinetic energy cascade and total flux.
The correlations of $\SE$ for the box filter are 
(within the error bars) slightly lower ($\lesssim 10\%$) for all models 
and filter widths.
In addition, the correlations exhibit a more pronounced deviation 
for the total energy flux $\FE$ especially at smaller filter wavenumbers $k$
and thus larger filter widths.
We attribute this to the non-smooth nature of the box kernel versus the 
Gaussian kernel resulting in 
numerical biases in the computation of gradient-based quantities.
This could explain why the deviations are more pronounced in the total flux
that has an additional divergence operator acting on the SGS terms in
comparison to the cascade flux.
Likewise, the effect would be more pronounced in the nonlinear closures as
they are built from nonlinear combinations of gradients.
The observed convergence between box and Gaussian filtering with increasing
$k$ is also expected, because the differences between the kernels become less 
distinct for small widths.
Overall, the observed behavior based on Gaussian filtering, i.e. better
performance of the nonlinear closures over the scale-similarity and 
the eddy-dissipation family ones, 
also holds for filtering with a box kernel.
These trends similarly apply to the cross-helicity fluxes and other SGS terms, 
too.

\subsection{Functional analysis: average SGS dissipation}
\begin{figure*}[htbp]
	\begin{center}
        \includegraphics[width=\textwidth]{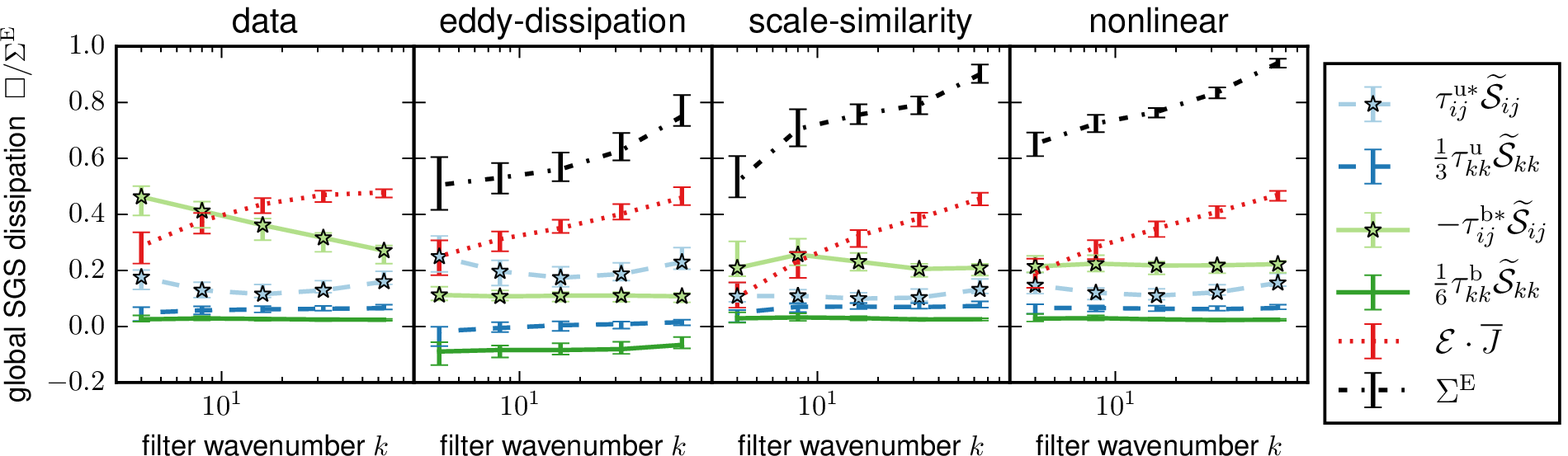}
	\end{center}
	\caption{Contributions of individual components 
		(deviatoric kinetic SGS stress, $\tuijdata[*]\Sflt_{ij}$, 
		and deviatoric magnetic SGS stress, $\tbijdata[*]\Sflt_{ij}$,
		kinetic SGS pressure, $1/3\tudata_{kk}\Sflt_{kk}$, 
		and magnetic SGS pressure, $1/6\tbdata_{kk}\Sflt_{kk}$,
		and EMF, $\EMFdata \cdot \Jflt$) normalized 
		to the average SGS dissipation, $\SE$,
		of supersonic simulation 7b for different filter widths. 
		The markers illustrate the median and the error bars show the minimum
		and maximum values over time.
		Each closure family is represented by the locally best performing closures, i.e.
		eddy-dissipation of $\EVE$-$\EkinsgsS$-$\EDW$-$\EmagsgsJ$-$\abg$,
		scale-similarity of $\SSu$-$\SSb$-$\SSemf$, and the nonlinear family
		of $\NLu$-$\NLb$-$\NLemfcompr$.
	}
	\label{fig:SGSdiss}
\end{figure*}
We close the functional analysis with a comparison of the contributions by 
individual components to the average SGS dissipation $\SE$.
Figure \ref{fig:SGSdiss} illustrates the share of
deviatoric kinetic SGS stress, $\tuijdata[*]\Sflt_{ij}$, 
and deviatoric magnetic SGS stress, $\tbijdata[*]\Sflt_{ij}$,
kinetic SGS pressure, $1/3\tudata_{kk}\Sflt_{kk}$, 
and magnetic SGS pressure, $1/6\tbdata_{kk}\Sflt_{kk}$,
and EMF, $\EMFdata \cdot \Jflt$ to $\SE$ for 
different filter widths.
In general, both SGS pressures (and thus energies) are almost negligible ($<10\%$) 
in the reference data even though the data covers the slightly supersonic regime 
(simulation 7b).
Similarly, the deviatoric kinetic SGS stress is subdominant ($10\%$-$20\%$) while
the deviatoric magnetic SGS stress and the EMF, which jointly contribute $\approx80\%$
to the total SGS dissipation independent of the chosen filter scale.
While the magnetic stress dominates at the largest scales (up to $50\%$ at $k=4$), its
contribution constantly decreases,
and at the smallest scale the EMF is strongest reaching a contribution
of $\approx 50\%$.
This can be understood by analyzing the ratio of forward to inverse energy transfer
(not shown).
While the forward transfer mediated by $\tbijdata[*]\Sflt_{ij}$ is $\approx30$ times 
stronger than the inverse transfer at $k=4$, it is only $\approx6$ times stronger at
$k=64$. At the same time the ratio by the EMF remains constantly at a factor $\approx 8$.
Two scenarios (or more likely an unbalanced combination thereof) 
could potentialy explain this situation:
either the existance of an inverse cascade coupled to direct forward transfer, or
direct inverse transfer coupled with a forward cascade.
On the one hand, a cascade typically transfers energy from one scale to the next smaller 
(or larger) scale resulting in a constant flux with varying filter width.
On the other hand, direct transfer allows exchange of energy between scales with arbitrary
separation and thus the flux may vary with varying filter width.
Although a  more detailed study, e.g. by a shell-to-shell energy transfer analysis, would allow 
a better interpretation, it is not required for the following closure discussion and we leave it as
subject to future work.

Before analyzing the predicted contributions by the different closure families,
it should be noted that the coefficient from the fit has been used to 
calculate the resulting dissipation values.
Allowing all coefficients to vary freely and optimizing for average SGS dissipation
would allow each closure to excatly match the reference data and, in turn, render
this analysis meaningless.
In general, all closure families behave similar with respect to the total dissipation.
At large scales they underestimate the reference data by $\approx50\%$ (eddy-dissipation
and scale-similarity) and $\approx40\%$ (nonlinear), while improving towards the 
smallest scales reaching $\approx75\%$ ($\ED$), $\approx90\%$ ($\SSid$) and 
$\approx95\%$ ($\NL$) agreement.
This is seen to be due to the successful capture of the EMF related contribution
and failing to represent the deviatoric magnetic stress dynamics at varying
filter scale.
In other words, all closures predict too much net inverse energy transfer to 
the largest scales.
Another important observation concerns the overall inverse energy transfer by
the magnetic SGS pressure of the eddy-diffusivity closure.
Given that the eddy-viscosity and eddy-resistivity closures can not provide inverse
energy transfer by construction, and that the eddy-diffusivity closure itself exhibits
the overall poorest correlation as shown in the previous subsections, the SGS pressures
are the only channels left for inverse transfer in this closure set.
Thus, in the process of matching the inverse transfer that is present in the reference data,
an over-compensation in the SGS energies takes place.

\subsection{Structural analysis: topology}
\begin{figure*}[htbp]
\centering
		\includegraphics{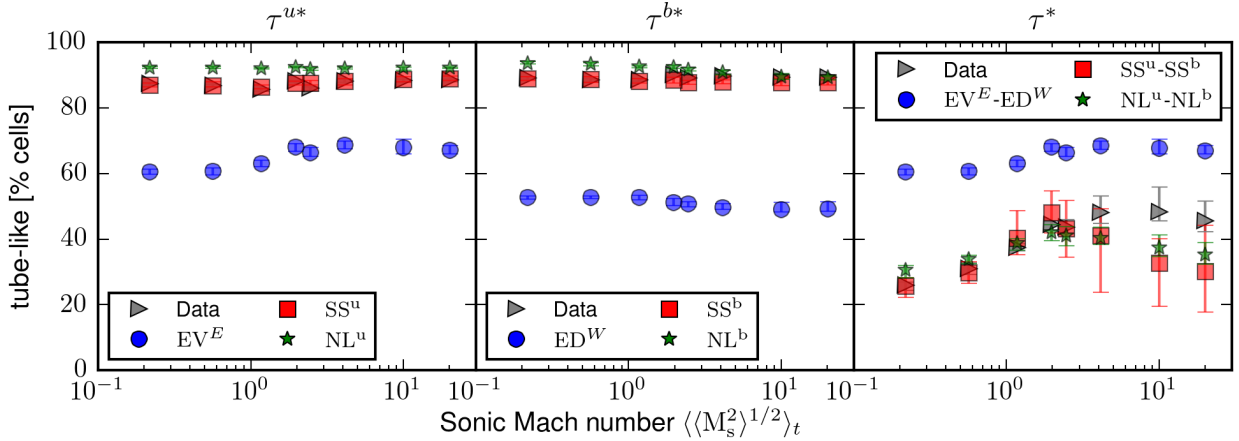}
	\caption{
Topology of deviatoric stress tensors by mean percentage of tube-like structures 
over all snapshots of each simulation (1, 2a, 4, 6, 7a, 11, 12 and 13,
see table \ref{tab:SimOverview}).
The remaining structures are sheet-like.
The error bars indicate the minimum and maximum value over time for each 
simulation.
}
	\label{fig:QR}
\end{figure*}
We begin our structural analysis with the comparison of the deviatoric stress
tensor topology.
Figure~\ref{fig:QR} illustrates the amount of tube-like structures in
our simulations.
The only other possibility for $\tudata[*]$, $\tbdata[*]$ and $\tau^*$
are sheet-like structures.
Analyzing the kinetic $\tudata[*]$ and magnetic $\tbdata[*]$ tensors
individually we have $\approx 88\%$ tube-like structures and $\approx 12\%$
sheet-like structures in the data independent of tensor and sonic Mach number
$\Ms$.
Furthermore, there are almost no temporal variations within each simulation
-- the error bars
indicating the minimum and maximum are within the markers.
The scale-similarity closures $\SSu$ and $\SSb$ match these topologies very 
closely with differences of only $\approx1\%$.
The nonlinear closures $\NLu$ and $\NLb$ are closely following the data 
topology as well, even though they slightly overestimate the amount of 
tube-like structure by $\approx 3\%$ in general.
Eddy-viscosity $\EVE$ and eddy-diffusivity $\EDW$ closures on the other hand 
are not able to match the flow topology.
While $\EVE$ is able to reproduce at least the correct tendency with dominating 
tube structures ($65\%$), $\EDW$ produces an equal share of
tube and sheet structures.
Interestingly, the topological configuration changes dramatically when
analyzing the deviatoric tensor $\tau^* = \tudata[*] - \tbdata[*]$
as a whole.
The dominant, $\Ms$-independent tube-like topology vanishes and 
sheet configurations become dominant in the subsonic regime.
In the supersonic regime tube- and sheet-like configurations are
equally present with some (\textless$10\%$) temporal variation.
Again, scale-similarity and nonlinear closures are able to follow
the trend more closely than eddy-dissipation type closures.
$\EVE$-$\EDW$ exhibits the same behavior as $\EVE$ alone and
provides mainly tube-like topology.
The scale-similarity closure correctly captures the topology in the
subsonic regime with negligible temporal variations.
However, in the supersonic regime the amount of sheet-like structures
is overestimated by $15\%$ on average and there are temporal
variations of up to $14\%$. 
In contrast to this, the nonlinear closure shows less variations 
(\textless$4\%$).
However, it also overestimates sheet-like structure in the 
supersonic regime, but by only $10\%$.
Overall these results are in line with the original closure 
approaches -- functional versus structural.
The functional eddy-dissipation closures do not perform well
in this structural test, whereas both structural closure 
families are capable of capturing the data topology.

\subsection{Structural analysis: alignment and magnitude}
In order to asses how the different closures perform as vectors 
in the equations, i.e. $\nabla \cdot \widehat\tau$ and $\nabla \times \EMF$, we
compare their magnitude and alignment with the reference data.
\begin{figure}[htbp]
\centering
		\includegraphics{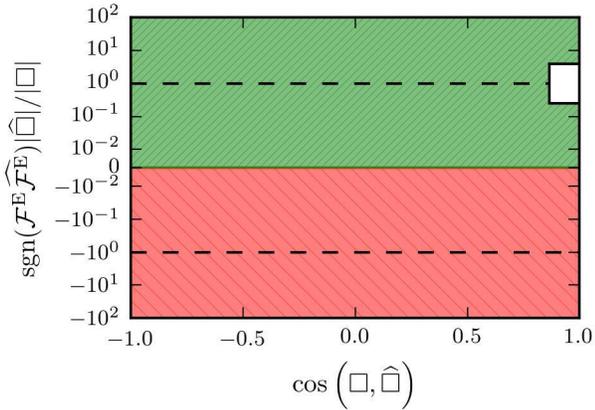}
	\caption{
Illustration of magnitude-alignment 2D-histograms (see figure
\ref{fig:alignments}). 
The x-axis shows the alignment between closure vector and reference vector.
Relative closure magnitudes are given on the y-axis with the dashed (- -) 
lines indicating identical closure and reference magnitude.
The upper half (green ``/'' hatching) indicates equal direction in
energy cascade, i.e. same sign of $\FE$ and $\widehat{\FE}$, 
whereas the lower half (red ``\textbackslash'' hatching)
corresponds to opposed directions. 
The white box illustrates the area of optimal performance:
alignment is within $30^\circ$, relative magnitude within a factor of 4, 
and identical flux sign.
}
\label{fig:align-expl}
\end{figure}
Figure~\ref{fig:align-expl} is an explanatory sketch of the 2D-histograms 
we use for the analysis.
The relative vector magnitude, e.g.  
$|\nabla \cdot \tu[*]|/|\nabla \cdot \tudata[*]|$,
is plotted against the angle between closure and exact solution, e.g.
$\cos \left ( \nabla \cdot \tu[*], \nabla \cdot \tudata[*] \right )$.
Furthermore, we use the sign of the product  of closure flux $\widehat{\FE}$  
and reference flux $\FE$ to split the histogram in two halves.
A positive sign corresponds to the right direction of the cascade, 
while a negative one indicates opposite direction.
We choose this kind of presentation as it illustrates several independent 
measures for single-coefficient closures.
Firstly, the magnitude is a direct result of the free coefficient value 
that is determined by the fitting process.
Secondly, the sign of the fluxes is determined in conjunction with a 
resolved flow quantity, e.g. $\fav{\VEC{u}}$ for 
$\FE = \fav{\VEC{u}} \cdot \bra{\nabla \cdot \tau}$, see \eqref{eq:FE}, and is independent of 
the coefficient magnitude.
Thirdly, the angle is given by the SGS terms alone and is also independent of the 
coefficient magnitude.
We define a region of optimal performance in order to make quantitative 
statements.
Within this region the relative magnitude does not deviate by
more than a factor of 4, the angle between closure and data is \textless$30^\circ$,
and both fluxes ($\widehat{\FE}$ and $\FE$) have identical sign.
We use the results of the energy flux fits $\FE$ in this subsection.
Nevertheless, we also verified that the conclusions similarly apply to the other 
flux fits $\SE$, $\SW$ and $\FW$.

\begin{figure*}[htbp]
\centering
		\subfigure{
		\includegraphics{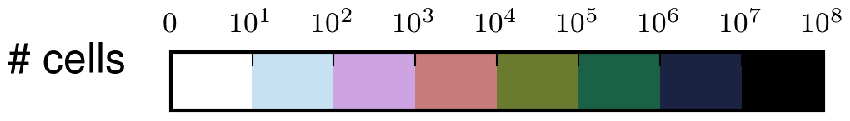}}
\setcounter{subfigure}{0}
		\subfigure[\label{fig:align-tauUStar}
Deviatoric SGS Reynolds stress closures {\protect $\tu[*]$}]{
		\includegraphics{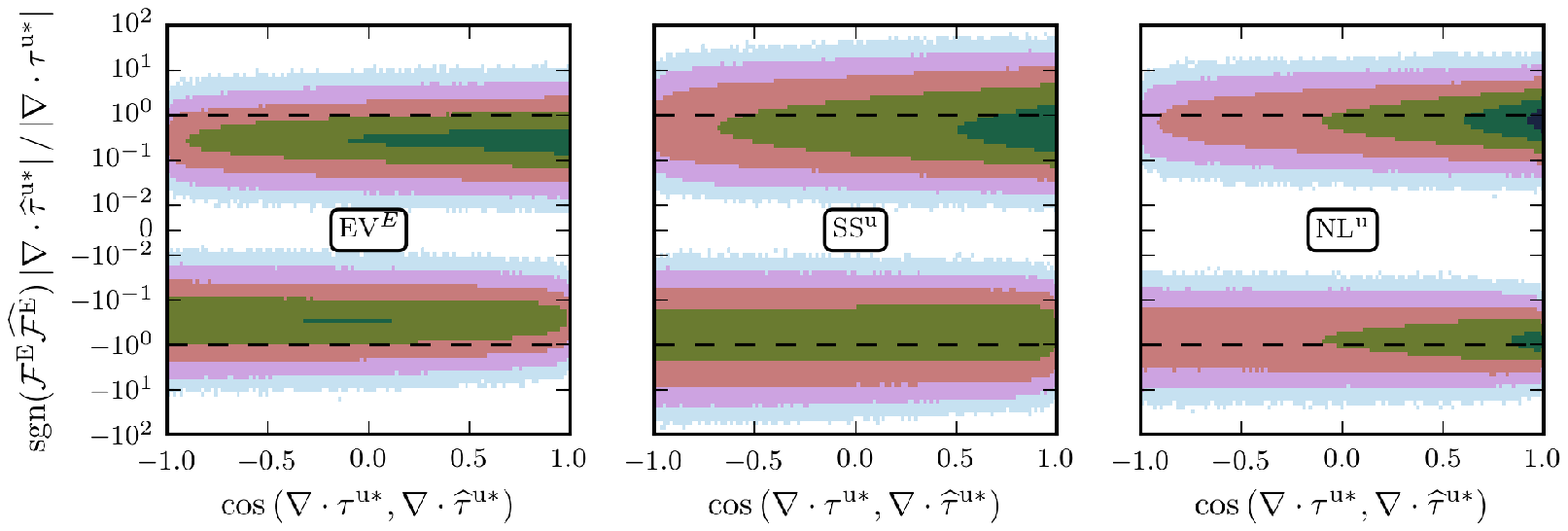}}
		\subfigure[\label{fig:align-tauBStar}
Deviatoric SGS Maxwell stress closures {\protect $\tb[*]$}]{
		\includegraphics{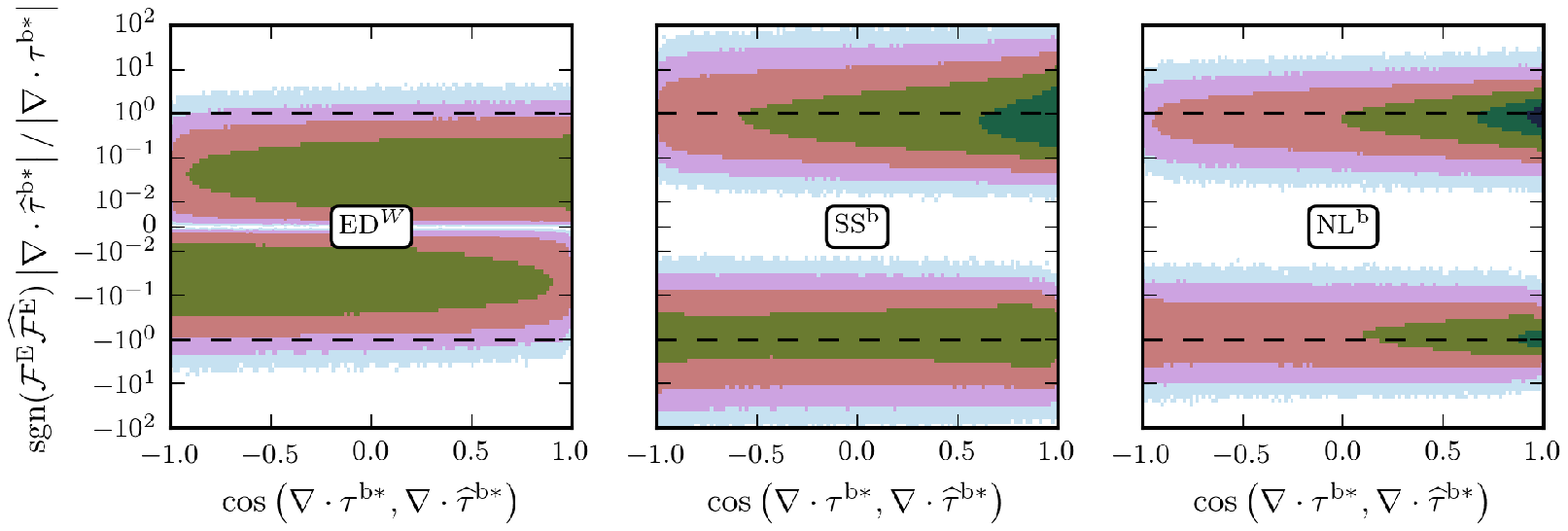}}
			\subfigure[\label{fig:align-EMF}
Electromotive force closures $\EMF$]{
		\includegraphics{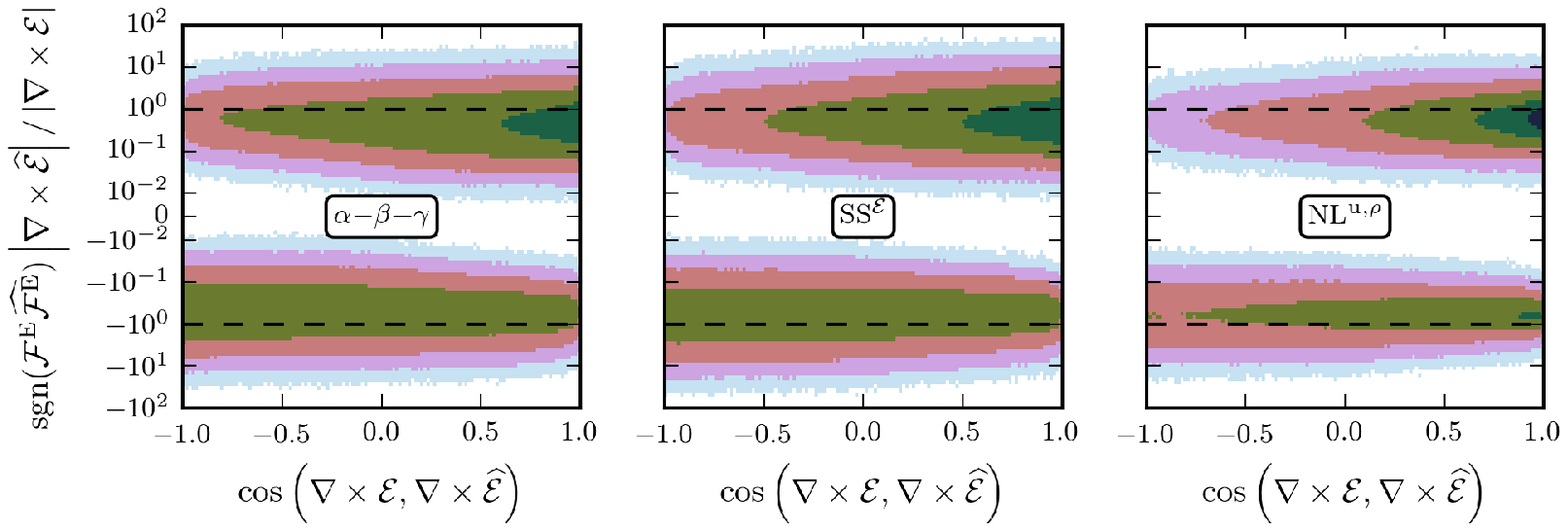}}
	\caption{Two dimensional histograms showing the distribution of relative closure
		vector magnitude, i.e.
$\sgn\bra{\FE \widehat \FE} |\nabla \cdot \tunoij[*]|/|\nabla \cdot \tudata[*]|$,
		versus alignment, i.e.
$\cos \left ( \nabla \cdot \tunoij[*], \nabla \cdot \tudata[*] \right )$,
	between closure and data vector.
The additional signum, $\sgn$, function on the y-axis is used to indicate
flux alignment, i.e. whether data flux $\FE$ and the flux predicted by the
closure $\widehat \FE$ have identical sign.
Dashed lines in each plot illustrate identical closure and data vector magnitudes.
The data is taken from a single snapshot at $t=4T$ of supersonic simulation 7a
	filtered at $k=16$.
}
	\label{fig:alignments}
\end{figure*}

Figure~\ref{fig:alignments} illustrates the resulting 2D-histograms for the best
performing closures in a snapshot of the supersonic simulation 7a at $t = 4T$,
which has randomly been chosen for illustration purposes.
The deviatoric SGS Reynolds stress $\tu[*]$ closures $\EVE$, $\SSu$ and $\NLu$
are shown in figure~\ref{fig:align-tauUStar}.
In general, the magnitude predicted by $\EVE$ and $\SSu$ is too small.
Furthermore, the angle between closure and data is almost randomly distributed
with a slight tendency of alignment, which is more pronounced for $\SSu$.
In contrast to this, $\NLu$ exhibits a clear peak at exact alignment and
equal magnitude.

Over all simulations $49^{+10}_{-4}\%$ (median and bounds giving
the maximum and minimum) of the cells within the simulation cube are
within the region of optimal performance for $\NLu$ and $81^{+3}_{-2}\%$
have the correct sign of $\FE$.
$\SSu$ has still $66^{+2}_{-2}\%$ cells with the correct sign and 
$14^{+4}_{-3}\%$ in the optimal region, whereas $\EVE$ performs worst
with $5^{+4}_{-4}\%$ in the optimal region and only $58^{+2}_{-2}\%$
with equal sign.

Figure~\ref{fig:align-tauBStar} illustrates the deviatoric SGS Maxwell 
closures $\NLb$, $\SSb$ and $\EDW$ for the same snapshot.
Overall, the nonlinear and scale-similarity closure behave very similar to their
kinetic counterparts with $61^{+13}_{-12}\%$ optimal region and 
$84^{+5}_{-5}\%$ correct sign for $\NLb$, and $27^{+5}_{-8}\%$ 
optimal region and $71^{+3}_{-2}\%$ correct sign for
$\SSb$, respectively.
The weak performance of eddy-diffusivity 
closures described in the previous section is also apparent here.
The magnitude of $\EDW$ is typically too small by more than a factor of 10.
This comes as no surprise as it is determined by the 
free coefficient.
Given that $\FE$ and $\widehat{\FE}$ have matching signs only 
in $52^{+1}_{-1}\%$ of the cells, which corresponds to
random behavior, the
fitting process favors a closure close to $0$.
In addition to this, the distribution of the angle between closure
and data, which is independent of the fitting procedure, is 
completely random and \textless$1\permil$ are in the optimal region.

Finally, the EMF closures $\abg$, $\SSemf$ and
$\NLemfcompr$ are depicted in figure~\ref{fig:align-EMF} for the same snapshot.
Here, the performance of the eddy-dissipation family closure
$\abg$ is best compared to the other terms.
Overall, $13^{+4}_{-4}\%$ cells are within the optimal region and
$61^{+5}_{-4}\%$ have the correct sign.
$\SSemf$ performs slightly better with $19^{+4}_{-7}\%$
and $66^{+2}_{-5}\%$, respectively.
In both cases the closure vector is
more likely to be aligned with the data vector even though
it is not as pronounced as for the $\NLemfcompr$ closure.
For the nonlinear closure $53^{+6}_{-29}\%$ are within the optimal region
whereby the lower limit stems from the highly supersonic simulations
12 and 13.
Nevertheless, $\NLemfcompr$ produces the correct flux sign in the 
majority of cells ($80^{+3}_{-8}\%$) and the variation is less extensive.

The general trend that nonlinear closures are performing best, followed
by scale-similarity closures and eventually eddy-dissipation closures
is again visible for all terms, $\tudata[*]$, $\tbdata[*]$ and $\EMFdata$.
\section{Conclusions and outlook}
\label{sec:conclusions}
In this paper we systematically conducted \aprio tests of different
subgrid-scale closures in the realm of compressible magnetohydrodynamics.
Over a large parameter space of 15 simulations of forced, homogeneous, isotropic
turbulence with sonic Mach numbers ranging from $\Ms = 0.2$ to $20$ we
were able to show that closures of the proposed nonlinear type outperform
traditional closures of eddy-dissipation and scale-similarity type in every
single test.
The main feature of the nonlinear closures is that they require no assumptions
about the nature of the flow or turbulence, and, therefore, are able to 
capture anisotropic effects and support up- and down-scale energy transfer.
In contrast, the scale-similarity and eddy-dissipation type closures assume some
universal behavior of turbulence.
The \aprio tests included the correlation between closure and explicitly 
filtered reference data for quantities such 
as the turbulent energy $\SE$ and cross-helicity $\SW$ cascades, 
and total turbulent energy $\FE$ and cross-helicity $\FW$ fluxes.
The turbulent energy cascade flux has also been used to analyze the
average SGS dissipation.
Additionally, we also evaluated the distribution of  topological structures 
for the SGS 
Reynolds and Maxwell stress tensors and their alignment with respect to the 
reference data in physical space.
Moreover, we verified that our conclusions are not sensitive to resolution,
filter width or filter kernel by
comparing results between $512^3$ and $1024^3$ resolution simulations at filter widths of 
$k=4,8,16,32,64$ with box kernel and a Gaussian kernel.
Finally, we were able to verify that the free coefficients of the basic 
nonlinear closures are very close to unity as expected from the
analytic derivation.

Overall, we conclude that the eddy-dissipation family including the 
popular Smagorinsky closure has only a limited range of applicability,
e.g. in situations with dominantly
supersonic turbulence and in situations where local flow features
are less important.
Closures of the scale-similarity family or the nonlinear family can be applied 
in much more diverse situations, e.g. where anisotropic features or 
up-scale energy transfer are required.
However, there is still room for improvement as the net up-scale transfer
via the SGS Maxwell stress is overestimated.
Furthermore, the scale-similarity closures should be handled with care as their 
performance varies strongly with reference quantity and sonic Mach number.
The basic nonlinear closures, $\NLu$, $\NLb$ and $\NLemfcompr$,  on the other
hand perform well across the entire parameter space and are able to reproduce
local flow features.

This encourages the application of the basic nonlinear closures
as a zero-coefficient SGS model
in large-eddy simulations of compressible MHD.
These simulations would benefit from the additional physics
provided by the SGS model.
Promising processes for such LES are turbulent
magnetic reconnection\cite{2015ApJ...806L..12O} or
the turbulent dynamo\cite{PhysRevE.92.023010}, for example, in 
star-forming magnetized clouds\cite{2004RvMP...76..125M}
or even in galaxies\cite{2041-8205-783-1-L20} and clusters.

\begin{acknowledgments}
The authors would like to thank C. Federrath for providing the \FLASH simulations.
  PG acknowledges financial support by the 
  \textit{International Max Planck Research School for Solar System 
  Science at the University of Göttingen}.
  DV acknowledge researchs funding by the 
  \textit{Deutsche Forschungsgemeinschaft (DFG)} under grant
  \textit{SFB 963/1, project A15}.
  DRGS thanks for funding through Fondecyt regular (project code 1161247) and 
  through the ''Concurso Proyectos Internacionales de Investigaci\'on, 
  Convocatoria 2015'' (project code PII20150171).
  The \Enzo simulations were performed and analyzed with the HLRN-III facilities of 
  the \textit{North-German Supercomputing Alliance} under grant \textit{nip00037}.

\end{acknowledgments}
\section*{References}


\begin{thebibliography}{50}%
\makeatletter
\providecommand \@ifxundefined [1]{%
 \@ifx{#1\undefined}
}%
\providecommand \@ifnum [1]{%
 \ifnum #1\expandafter \@firstoftwo
 \else \expandafter \@secondoftwo
 \fi
}%
\providecommand \@ifx [1]{%
 \ifx #1\expandafter \@firstoftwo
 \else \expandafter \@secondoftwo
 \fi
}%
\providecommand \natexlab [1]{#1}%
\providecommand \enquote  [1]{``#1''}%
\providecommand \bibnamefont  [1]{#1}%
\providecommand \bibfnamefont [1]{#1}%
\providecommand \citenamefont [1]{#1}%
\providecommand \href@noop [0]{\@secondoftwo}%
\providecommand \href [0]{\begingroup \@sanitize@url \@href}%
\providecommand \@href[1]{\@@startlink{#1}\@@href}%
\providecommand \@@href[1]{\endgroup#1\@@endlink}%
\providecommand \@sanitize@url [0]{\catcode `\\12\catcode `\$12\catcode
  `\&12\catcode `\#12\catcode `\^12\catcode `\_12\catcode `\%12\relax}%
\providecommand \@@startlink[1]{}%
\providecommand \@@endlink[0]{}%
\providecommand \url  [0]{\begingroup\@sanitize@url \@url }%
\providecommand \@url [1]{\endgroup\@href {#1}{\urlprefix }}%
\providecommand \urlprefix  [0]{URL }%
\providecommand \Eprint [0]{\href }%
\providecommand \doibase [0]{http://dx.doi.org/}%
\providecommand \selectlanguage [0]{\@gobble}%
\providecommand \bibinfo  [0]{\@secondoftwo}%
\providecommand \bibfield  [0]{\@secondoftwo}%
\providecommand \translation [1]{[#1]}%
\providecommand \BibitemOpen [0]{}%
\providecommand \bibitemStop [0]{}%
\providecommand \bibitemNoStop [0]{.\EOS\space}%
\providecommand \EOS [0]{\spacefactor3000\relax}%
\providecommand \BibitemShut  [1]{\csname bibitem#1\endcsname}%
\let\auto@bib@innerbib\@empty
\bibitem [{\citenamefont {{Cooper}}\ \emph {et~al.}(2014)\citenamefont
  {{Cooper}}, \citenamefont {{Wallace}}, \citenamefont {{Brookhart}},
  \citenamefont {{Clark}}, \citenamefont {{Collins}}, \citenamefont {{Ding}},
  \citenamefont {{Flanagan}}, \citenamefont {{Khalzov}}, \citenamefont {{Li}},
  \citenamefont {{Milhone}}, \citenamefont {{Nornberg}}, \citenamefont
  {{Nonn}}, \citenamefont {{Weisberg}}, \citenamefont {{Whyte}}, \citenamefont
  {{Zweibel}},\ and\ \citenamefont {{Forest}}}]{2014PhPl...21a3505C}%
  \BibitemOpen
  \bibfield  {author} {\bibinfo {author} {\bibfnamefont {C.~M.}\ \bibnamefont
  {{Cooper}}}, \bibinfo {author} {\bibfnamefont {J.}~\bibnamefont {{Wallace}}},
  \bibinfo {author} {\bibfnamefont {M.}~\bibnamefont {{Brookhart}}}, \bibinfo
  {author} {\bibfnamefont {M.}~\bibnamefont {{Clark}}}, \bibinfo {author}
  {\bibfnamefont {C.}~\bibnamefont {{Collins}}}, \bibinfo {author}
  {\bibfnamefont {W.~X.}\ \bibnamefont {{Ding}}}, \bibinfo {author}
  {\bibfnamefont {K.}~\bibnamefont {{Flanagan}}}, \bibinfo {author}
  {\bibfnamefont {I.}~\bibnamefont {{Khalzov}}}, \bibinfo {author}
  {\bibfnamefont {Y.}~\bibnamefont {{Li}}}, \bibinfo {author} {\bibfnamefont
  {J.}~\bibnamefont {{Milhone}}}, \bibinfo {author} {\bibfnamefont
  {M.}~\bibnamefont {{Nornberg}}}, \bibinfo {author} {\bibfnamefont
  {P.}~\bibnamefont {{Nonn}}}, \bibinfo {author} {\bibfnamefont
  {D.}~\bibnamefont {{Weisberg}}}, \bibinfo {author} {\bibfnamefont {D.~G.}\
  \bibnamefont {{Whyte}}}, \bibinfo {author} {\bibfnamefont {E.}~\bibnamefont
  {{Zweibel}}}, \ and\ \bibinfo {author} {\bibfnamefont {C.~B.}\ \bibnamefont
  {{Forest}}},\ }\href {\doibase 10.1063/1.4861609} {\bibfield  {journal}
  {\bibinfo  {journal} {Physics of Plasmas}\ }\textbf {\bibinfo {volume}
  {21}},\ \bibinfo {eid} {013505} (\bibinfo {year} {2014})}\BibitemShut
  {NoStop}%
\bibitem [{\citenamefont {{Oishi}}\ \emph {et~al.}(2015)\citenamefont
  {{Oishi}}, \citenamefont {{Mac Low}}, \citenamefont {{Collins}},\ and\
  \citenamefont {{Tamura}}}]{2015ApJ...806L..12O}%
  \BibitemOpen
  \bibfield  {author} {\bibinfo {author} {\bibfnamefont {J.~S.}\ \bibnamefont
  {{Oishi}}}, \bibinfo {author} {\bibfnamefont {M.-M.}\ \bibnamefont {{Mac
  Low}}}, \bibinfo {author} {\bibfnamefont {D.~C.}\ \bibnamefont {{Collins}}},
  \ and\ \bibinfo {author} {\bibfnamefont {M.}~\bibnamefont {{Tamura}}},\
  }\href {\doibase 10.1088/2041-8205/806/1/L12} {\bibfield  {journal} {\bibinfo
   {journal} {\apjl}\ }\textbf {\bibinfo {volume} {806}},\ \bibinfo {eid} {L12}
  (\bibinfo {year} {2015})},\ \Eprint {http://arxiv.org/abs/1505.04653}
  {arXiv:1505.04653 [astro-ph.SR]} \BibitemShut {NoStop}%
\bibitem [{\citenamefont {Schober}\ \emph {et~al.}(2015)\citenamefont
  {Schober}, \citenamefont {Schleicher}, \citenamefont {Federrath},
  \citenamefont {Bovino},\ and\ \citenamefont {Klessen}}]{PhysRevE.92.023010}%
  \BibitemOpen
  \bibfield  {author} {\bibinfo {author} {\bibfnamefont {J.}~\bibnamefont
  {Schober}}, \bibinfo {author} {\bibfnamefont {D.~R.~G.}\ \bibnamefont
  {Schleicher}}, \bibinfo {author} {\bibfnamefont {C.}~\bibnamefont
  {Federrath}}, \bibinfo {author} {\bibfnamefont {S.}~\bibnamefont {Bovino}}, \
  and\ \bibinfo {author} {\bibfnamefont {R.~S.}\ \bibnamefont {Klessen}},\
  }\href {\doibase 10.1103/PhysRevE.92.023010} {\bibfield  {journal} {\bibinfo
  {journal} {Phys. Rev. E}\ }\textbf {\bibinfo {volume} {92}},\ \bibinfo
  {pages} {023010} (\bibinfo {year} {2015})}\BibitemShut {NoStop}%
\bibitem [{\citenamefont {{Goldstein}}, \citenamefont {{Roberts}},\ and\
  \citenamefont {{Matthaeus}}(1995)}]{Goldstein1995}%
  \BibitemOpen
  \bibfield  {author} {\bibinfo {author} {\bibfnamefont {M.~L.}\ \bibnamefont
  {{Goldstein}}}, \bibinfo {author} {\bibfnamefont {D.~A.}\ \bibnamefont
  {{Roberts}}}, \ and\ \bibinfo {author} {\bibfnamefont {W.~H.}\ \bibnamefont
  {{Matthaeus}}},\ }\href {\doibase 10.1146/annurev.aa.33.090195.001435}
  {\bibfield  {journal} {\bibinfo  {journal} {\araa}\ }\textbf {\bibinfo
  {volume} {33}},\ \bibinfo {pages} {283} (\bibinfo {year} {1995})}\BibitemShut
  {NoStop}%
\bibitem [{\citenamefont {Balbus}\ and\ \citenamefont
  {Hawley}(1998)}]{RevModPhys.70.1}%
  \BibitemOpen
  \bibfield  {author} {\bibinfo {author} {\bibfnamefont {S.~A.}\ \bibnamefont
  {Balbus}}\ and\ \bibinfo {author} {\bibfnamefont {J.~F.}\ \bibnamefont
  {Hawley}},\ }\href {\doibase 10.1103/RevModPhys.70.1} {\bibfield  {journal}
  {\bibinfo  {journal} {Rev. Mod. Phys.}\ }\textbf {\bibinfo {volume} {70}},\
  \bibinfo {pages} {1} (\bibinfo {year} {1998})}\BibitemShut {NoStop}%
\bibitem [{\citenamefont {Schmidt}(2015)}]{lrca-2015-2}%
  \BibitemOpen
  \bibfield  {author} {\bibinfo {author} {\bibfnamefont {W.}~\bibnamefont
  {Schmidt}},\ }\href {\doibase 10.1007/lrca-2015-2} {\bibfield  {journal}
  {\bibinfo  {journal} {Living Reviews in Computational Astrophysics}\ }\textbf
  {\bibinfo {volume} {1}} (\bibinfo {year} {2015}),\
  10.1007/lrca-2015-2}\BibitemShut {NoStop}%
\bibitem [{\citenamefont {Miesch}\ \emph {et~al.}(2015)\citenamefont {Miesch},
  \citenamefont {Matthaeus}, \citenamefont {Brandenburg}, \citenamefont
  {Petrosyan}, \citenamefont {Pouquet}, \citenamefont {Cambon}, \citenamefont
  {Jenko}, \citenamefont {Uzdensky}, \citenamefont {Stone}, \citenamefont
  {Tobias}, \citenamefont {Toomre},\ and\ \citenamefont {Velli}}]{Miesch2015}%
  \BibitemOpen
  \bibfield  {author} {\bibinfo {author} {\bibfnamefont {M.}~\bibnamefont
  {Miesch}}, \bibinfo {author} {\bibfnamefont {W.}~\bibnamefont {Matthaeus}},
  \bibinfo {author} {\bibfnamefont {A.}~\bibnamefont {Brandenburg}}, \bibinfo
  {author} {\bibfnamefont {A.}~\bibnamefont {Petrosyan}}, \bibinfo {author}
  {\bibfnamefont {A.}~\bibnamefont {Pouquet}}, \bibinfo {author} {\bibfnamefont
  {C.}~\bibnamefont {Cambon}}, \bibinfo {author} {\bibfnamefont
  {F.}~\bibnamefont {Jenko}}, \bibinfo {author} {\bibfnamefont
  {D.}~\bibnamefont {Uzdensky}}, \bibinfo {author} {\bibfnamefont
  {J.}~\bibnamefont {Stone}}, \bibinfo {author} {\bibfnamefont
  {S.}~\bibnamefont {Tobias}}, \bibinfo {author} {\bibfnamefont
  {J.}~\bibnamefont {Toomre}}, \ and\ \bibinfo {author} {\bibfnamefont
  {M.}~\bibnamefont {Velli}},\ }\href {\doibase 10.1007/s11214-015-0190-7}
  {\bibfield  {journal} {\bibinfo  {journal} {Space Science Reviews}\ ,\
  \bibinfo {pages} {1}} (\bibinfo {year} {2015})}\BibitemShut {NoStop}%
\bibitem [{\citenamefont {Chernyshov}, \citenamefont {Karelsky},\ and\
  \citenamefont {Petrosyan}(2014)}]{Chernyshov2014}%
  \BibitemOpen
  \bibfield  {author} {\bibinfo {author} {\bibfnamefont {A.~A.}\ \bibnamefont
  {Chernyshov}}, \bibinfo {author} {\bibfnamefont {K.~V.}\ \bibnamefont
  {Karelsky}}, \ and\ \bibinfo {author} {\bibfnamefont {A.~S.}\ \bibnamefont
  {Petrosyan}},\ }\href {\doibase 10.3367/UFNe.0184.201405a.0457} {\bibfield
  {journal} {\bibinfo  {journal} {Physics-Uspekhi}\ }\textbf {\bibinfo {volume}
  {57}},\ \bibinfo {pages} {421} (\bibinfo {year} {2014})}\BibitemShut
  {NoStop}%
\bibitem [{\citenamefont {Vlaykov}(2015)}]{Vlaykov15}%
  \BibitemOpen
  \bibfield  {author} {\bibinfo {author} {\bibfnamefont {D.~G.}\ \bibnamefont
  {Vlaykov}},\ }\emph {\bibinfo {title} {Sub-grid Scale Modelling of
  Compressible Magnetohydrodynamic Turbulence: Derivation and A Priori
  Analysis}},\ \href {http://hdl.handle.net/11858/00-1735-0000-0028-866C-C}
  {Ph.D. thesis},\ \bibinfo  {school} {Georg-August University School of
  Science (GAUSS) Göttingen} (\bibinfo {year} {2015})\BibitemShut {NoStop}%
\bibitem [{\citenamefont {{Favre}}(1983)}]{1983PhFl...26.2851F}%
  \BibitemOpen
  \bibfield  {author} {\bibinfo {author} {\bibfnamefont {A.}~\bibnamefont
  {{Favre}}},\ }\href {\doibase 10.1063/1.864049} {\bibfield  {journal}
  {\bibinfo  {journal} {Physics of Fluids}\ }\textbf {\bibinfo {volume} {26}},\
  \bibinfo {pages} {2851} (\bibinfo {year} {1983})}\BibitemShut {NoStop}%
\bibitem [{\citenamefont {Sagaut}(2006)}]{Sagaut2006}%
  \BibitemOpen
  \bibfield  {author} {\bibinfo {author} {\bibfnamefont {P.}~\bibnamefont
  {Sagaut}},\ }\href
  {http://www.springer.com/de/book/9783540263449?wt_mc=ThirdParty.SpringerLink.3.EPR653.About_eBook}
  {\emph {\bibinfo {title} {Large Eddy Simulation for Incompressible Flows: An
  Introduction}}},\ Scientific Computation\ (\bibinfo  {publisher} {Springer},\
  \bibinfo {year} {2006})\BibitemShut {NoStop}%
\bibitem [{\citenamefont {Lu}\ and\ \citenamefont
  {Porté-Agel}(2010)}]{LuPorte-Agel10}%
  \BibitemOpen
  \bibfield  {author} {\bibinfo {author} {\bibfnamefont {H.}~\bibnamefont
  {Lu}}\ and\ \bibinfo {author} {\bibfnamefont {F.}~\bibnamefont
  {Porté-Agel}},\ }\href {\doibase http://dx.doi.org/10.1063/1.3291073}
  {\bibfield  {journal} {\bibinfo  {journal} {Physics of Fluids}\ }\textbf
  {\bibinfo {volume} {22}},\ \bibinfo {eid} {015109} (\bibinfo {year} {2010}),\
  http://dx.doi.org/10.1063/1.3291073}\BibitemShut {NoStop}%
\bibitem [{\citenamefont {VREMAN}, \citenamefont {GEURTS},\ and\ \citenamefont
  {KUERTEN}(1997)}]{FLM:13495}%
  \BibitemOpen
  \bibfield  {author} {\bibinfo {author} {\bibfnamefont {B.}~\bibnamefont
  {VREMAN}}, \bibinfo {author} {\bibfnamefont {B.}~\bibnamefont {GEURTS}}, \
  and\ \bibinfo {author} {\bibfnamefont {H.}~\bibnamefont {KUERTEN}},\ }\href
  {\doibase 10.1017/S0022112097005429} {\bibfield  {journal} {\bibinfo
  {journal} {Journal of Fluid Mechanics}\ }\textbf {\bibinfo {volume} {339}},\
  \bibinfo {pages} {357} (\bibinfo {year} {1997})}\BibitemShut {NoStop}%
\bibitem [{\citenamefont {Balarac}\ \emph {et~al.}(2013)\citenamefont
  {Balarac}, \citenamefont {Le~Sommer}, \citenamefont {Meunier},\ and\
  \citenamefont {Vollant}}]{Balarac2013}%
  \BibitemOpen
  \bibfield  {author} {\bibinfo {author} {\bibfnamefont {G.}~\bibnamefont
  {Balarac}}, \bibinfo {author} {\bibfnamefont {J.}~\bibnamefont {Le~Sommer}},
  \bibinfo {author} {\bibfnamefont {X.}~\bibnamefont {Meunier}}, \ and\
  \bibinfo {author} {\bibfnamefont {A.}~\bibnamefont {Vollant}},\ }\href
  {\doibase http://dx.doi.org/10.1063/1.4813812} {\bibfield  {journal}
  {\bibinfo  {journal} {Physics of Fluids}\ }\textbf {\bibinfo {volume} {25}},\
  \bibinfo {eid} {075107} (\bibinfo {year} {2013}),\
  http://dx.doi.org/10.1063/1.4813812}\BibitemShut {NoStop}%
\bibitem [{\citenamefont {Braun}\ \emph {et~al.}(2014)\citenamefont {Braun},
  \citenamefont {Schmidt}, \citenamefont {Niemeyer},\ and\ \citenamefont
  {Almgren}}]{Braun21082014}%
  \BibitemOpen
  \bibfield  {author} {\bibinfo {author} {\bibfnamefont {H.}~\bibnamefont
  {Braun}}, \bibinfo {author} {\bibfnamefont {W.}~\bibnamefont {Schmidt}},
  \bibinfo {author} {\bibfnamefont {J.~C.}\ \bibnamefont {Niemeyer}}, \ and\
  \bibinfo {author} {\bibfnamefont {A.~S.}\ \bibnamefont {Almgren}},\ }\href
  {\doibase 10.1093/mnras/stu1119} {\bibfield  {journal} {\bibinfo  {journal}
  {Monthly Notices of the Royal Astronomical Society}\ }\textbf {\bibinfo
  {volume} {442}},\ \bibinfo {pages} {3407} (\bibinfo {year}
  {2014})}\BibitemShut {NoStop}%
\bibitem [{\citenamefont {Latif}\ \emph {et~al.}(2013)\citenamefont {Latif},
  \citenamefont {Schleicher}, \citenamefont {Schmidt},\ and\ \citenamefont
  {Niemeyer}}]{2013MNRAS.436.2989L}%
  \BibitemOpen
  \bibfield  {author} {\bibinfo {author} {\bibfnamefont {M.~A.}\ \bibnamefont
  {Latif}}, \bibinfo {author} {\bibfnamefont {D.~R.~G.}\ \bibnamefont
  {Schleicher}}, \bibinfo {author} {\bibfnamefont {W.}~\bibnamefont {Schmidt}},
  \ and\ \bibinfo {author} {\bibfnamefont {J.~C.}\ \bibnamefont {Niemeyer}},\
  }\href {\doibase 10.1093/mnras/stt1786} {\bibfield  {journal} {\bibinfo
  {journal} {Monthly Notices of the Royal Astronomical Society}\ }\textbf
  {\bibinfo {volume} {436}},\ \bibinfo {pages} {2989} (\bibinfo {year}
  {2013})}\BibitemShut {NoStop}%
\bibitem [{\citenamefont {Miki}\ and\ \citenamefont {Menon}(2008)}]{Miki2008}%
  \BibitemOpen
  \bibfield  {author} {\bibinfo {author} {\bibfnamefont {K.}~\bibnamefont
  {Miki}}\ and\ \bibinfo {author} {\bibfnamefont {S.}~\bibnamefont {Menon}},\
  }\href {\doibase http://dx.doi.org/10.1063/1.2947312} {\bibfield  {journal}
  {\bibinfo  {journal} {Physics of Plasmas (1994-present)}\ }\textbf {\bibinfo
  {volume} {15}},\ \bibinfo {eid} {072306} (\bibinfo {year}
  {2008})}\BibitemShut {NoStop}%
\bibitem [{\citenamefont {{Theobald}}, \citenamefont {{Fox}},\ and\
  \citenamefont {{Sofia}}(1994)}]{1994PhPl....1.3016T}%
  \BibitemOpen
  \bibfield  {author} {\bibinfo {author} {\bibfnamefont {M.~L.}\ \bibnamefont
  {{Theobald}}}, \bibinfo {author} {\bibfnamefont {P.~A.}\ \bibnamefont
  {{Fox}}}, \ and\ \bibinfo {author} {\bibfnamefont {S.}~\bibnamefont
  {{Sofia}}},\ }\href {\doibase 10.1063/1.870542} {\bibfield  {journal}
  {\bibinfo  {journal} {Physics of Plasmas}\ }\textbf {\bibinfo {volume} {1}},\
  \bibinfo {pages} {3016} (\bibinfo {year} {1994})}\BibitemShut {NoStop}%
\bibitem [{\citenamefont {Müller}\ and\ \citenamefont
  {Carati}(2002)}]{Mueller2002}%
  \BibitemOpen
  \bibfield  {author} {\bibinfo {author} {\bibfnamefont {W.-C.}\ \bibnamefont
  {Müller}}\ and\ \bibinfo {author} {\bibfnamefont {D.}~\bibnamefont
  {Carati}},\ }\href {\doibase http://dx.doi.org/10.1063/1.1448498} {\bibfield
  {journal} {\bibinfo  {journal} {Physics of Plasmas (1994-present)}\ }\textbf
  {\bibinfo {volume} {9}},\ \bibinfo {pages} {824} (\bibinfo {year}
  {2002})}\BibitemShut {NoStop}%
\bibitem [{\citenamefont {Yokoi}(2013)}]{Yokoi2013}%
  \BibitemOpen
  \bibfield  {author} {\bibinfo {author} {\bibfnamefont {N.}~\bibnamefont
  {Yokoi}},\ }\href
  {http://www.tandfonline.com/doi/abs/10.1080/03091929.2012.754022} {\bibfield
  {journal} {\bibinfo  {journal} {Geophys. Astrophys. Fluid Dyn.}\ ,\ \bibinfo
  {pages} {37}} (\bibinfo {year} {2013})}\BibitemShut {NoStop}%
\bibitem [{\citenamefont {{Balarac}}\ \emph {et~al.}(2010)\citenamefont
  {{Balarac}}, \citenamefont {{Kosovichev}}, \citenamefont {{Brugi{\`e}re}},
  \citenamefont {{Wray}},\ and\ \citenamefont {{Mansour}}}]{Balarac2010}%
  \BibitemOpen
  \bibfield  {author} {\bibinfo {author} {\bibfnamefont {G.}~\bibnamefont
  {{Balarac}}}, \bibinfo {author} {\bibfnamefont {A.~G.}\ \bibnamefont
  {{Kosovichev}}}, \bibinfo {author} {\bibfnamefont {O.}~\bibnamefont
  {{Brugi{\`e}re}}}, \bibinfo {author} {\bibfnamefont {A.~A.}\ \bibnamefont
  {{Wray}}}, \ and\ \bibinfo {author} {\bibfnamefont {N.~N.}\ \bibnamefont
  {{Mansour}}},\ }in\ \href
  {http://ctr.stanford.edu/Summer/SP10/8_06_balarac.pdf} {\emph {\bibinfo
  {booktitle} {{Proceedings of the Summer Program}}}}\ (\bibinfo {address}
  {Center for Turbulence Research, Stanford University/NASA},\ \bibinfo {year}
  {2010})\ pp.\ \bibinfo {pages} {503--512}\BibitemShut {NoStop}%
\bibitem [{\citenamefont {Grete}\ \emph {et~al.}(2015)\citenamefont {Grete},
  \citenamefont {Vlaykov}, \citenamefont {Schmidt}, \citenamefont
  {Schleicher},\ and\ \citenamefont {Federrath}}]{Grete2015}%
  \BibitemOpen
  \bibfield  {author} {\bibinfo {author} {\bibfnamefont {P.}~\bibnamefont
  {Grete}}, \bibinfo {author} {\bibfnamefont {D.~G.}\ \bibnamefont {Vlaykov}},
  \bibinfo {author} {\bibfnamefont {W.}~\bibnamefont {Schmidt}}, \bibinfo
  {author} {\bibfnamefont {D.~R.~G.}\ \bibnamefont {Schleicher}}, \ and\
  \bibinfo {author} {\bibfnamefont {C.}~\bibnamefont {Federrath}},\ }\href
  {http://stacks.iop.org/1367-2630/17/i=2/a=023070} {\bibfield  {journal}
  {\bibinfo  {journal} {New Journal of Physics}\ }\textbf {\bibinfo {volume}
  {17}},\ \bibinfo {pages} {023070} (\bibinfo {year} {2015})}\BibitemShut
  {NoStop}%
\bibitem [{\citenamefont {Vlaykov}\ \emph {et~al.}(2016)\citenamefont
  {Vlaykov}, \citenamefont {Grete}, \citenamefont {Schmidt},\ and\
  \citenamefont {Schleicher}}]{Vlaykov2016a}%
  \BibitemOpen
  \bibfield  {author} {\bibinfo {author} {\bibfnamefont {D.~G.}\ \bibnamefont
  {Vlaykov}}, \bibinfo {author} {\bibfnamefont {P.}~\bibnamefont {Grete}},
  \bibinfo {author} {\bibfnamefont {W.}~\bibnamefont {Schmidt}}, \ and\
  \bibinfo {author} {\bibfnamefont {D.~R.~G.}\ \bibnamefont {Schleicher}},\
  }\href@noop {} {\bibfield  {journal} {\bibinfo  {journal} {\textit{to be
  published in} Physics of Plasmas}\ } (\bibinfo {year} {2016})}\BibitemShut
  {NoStop}%
\bibitem [{\citenamefont {{Smagorinsky}}(1963)}]{Smagorinsky1963}%
  \BibitemOpen
  \bibfield  {author} {\bibinfo {author} {\bibfnamefont {J.}~\bibnamefont
  {{Smagorinsky}}},\ }\href {\doibase
  10.1175/1520-0493(1963)091<0099:GCEWTP>2.3.CO;2} {\bibfield  {journal}
  {\bibinfo  {journal} {Monthly Weather Review}\ }\textbf {\bibinfo {volume}
  {91}},\ \bibinfo {pages} {99} (\bibinfo {year} {1963})}\BibitemShut {NoStop}%
\bibitem [{\citenamefont {Vreman}, \citenamefont {Geurts},\ and\ \citenamefont
  {Kuerten}(1994)}]{Vreman1994}%
  \BibitemOpen
  \bibfield  {author} {\bibinfo {author} {\bibfnamefont {B.}~\bibnamefont
  {Vreman}}, \bibinfo {author} {\bibfnamefont {B.}~\bibnamefont {Geurts}}, \
  and\ \bibinfo {author} {\bibfnamefont {H.}~\bibnamefont {Kuerten}},\ }\href
  {\doibase 10.1017/S0022112094003745} {\bibfield  {journal} {\bibinfo
  {journal} {Journal of Fluid Mechanics}\ }\textbf {\bibinfo {volume} {278}},\
  \bibinfo {pages} {351} (\bibinfo {year} {1994})}\BibitemShut {NoStop}%
\bibitem [{\citenamefont {Agullo}\ \emph {et~al.}(2001)\citenamefont {Agullo},
  \citenamefont {Müller}, \citenamefont {Knaepen},\ and\ \citenamefont
  {Carati}}]{Agullo2001}%
  \BibitemOpen
  \bibfield  {author} {\bibinfo {author} {\bibfnamefont {O.}~\bibnamefont
  {Agullo}}, \bibinfo {author} {\bibfnamefont {W.-C.}\ \bibnamefont {Müller}},
  \bibinfo {author} {\bibfnamefont {B.}~\bibnamefont {Knaepen}}, \ and\
  \bibinfo {author} {\bibfnamefont {D.}~\bibnamefont {Carati}},\ }\href
  {\doibase http://dx.doi.org/10.1063/1.1372337} {\bibfield  {journal}
  {\bibinfo  {journal} {Physics of Plasmas}\ }\textbf {\bibinfo {volume} {8}},\
  \bibinfo {pages} {3502} (\bibinfo {year} {2001})}\BibitemShut {NoStop}%
\bibitem [{\citenamefont {{Widmer}}, \citenamefont {{B{\"u}chner}},\ and\
  \citenamefont {{Yokoi}}(2015)}]{2015arXiv151104347W}%
  \BibitemOpen
  \bibfield  {author} {\bibinfo {author} {\bibfnamefont {F.}~\bibnamefont
  {{Widmer}}}, \bibinfo {author} {\bibfnamefont {J.}~\bibnamefont
  {{B{\"u}chner}}}, \ and\ \bibinfo {author} {\bibfnamefont {N.}~\bibnamefont
  {{Yokoi}}},\ }\href@noop {} {\bibfield  {journal} {\bibinfo  {journal} {ArXiv
  e-prints}\ } (\bibinfo {year} {2015})},\ \Eprint
  {http://arxiv.org/abs/1511.04347} {arXiv:1511.04347 [physics.plasm-ph]}
  \BibitemShut {NoStop}%
\bibitem [{\citenamefont {{Bardina}}, \citenamefont {{Ferziger}},\ and\
  \citenamefont {{Reynolds}}(1980)}]{1980fpdy.confT....B}%
  \BibitemOpen
  \bibinfo {editor} {\bibfnamefont {J.}~\bibnamefont {{Bardina}}}, \bibinfo
  {editor} {\bibfnamefont {J.~H.}\ \bibnamefont {{Ferziger}}}, \ and\ \bibinfo
  {editor} {\bibfnamefont {W.~C.}\ \bibnamefont {{Reynolds}}},\ eds.,\
  \href@noop {} {\emph {\bibinfo {title} {American Institute of Aeronautics and
  Astronautics, Fluid and Plasma Dynamics Conference, 13th, Snowmass, Colo.,
  July 14-16, 1980, 10 p.}}}\ (\bibinfo {year} {1980})\BibitemShut {NoStop}%
\bibitem [{\citenamefont {Liu}, \citenamefont {Meneveau},\ and\ \citenamefont
  {Katz}(1994)}]{FLM:352690}%
  \BibitemOpen
  \bibfield  {author} {\bibinfo {author} {\bibfnamefont {S.}~\bibnamefont
  {Liu}}, \bibinfo {author} {\bibfnamefont {C.}~\bibnamefont {Meneveau}}, \
  and\ \bibinfo {author} {\bibfnamefont {J.}~\bibnamefont {Katz}},\ }\href
  {\doibase 10.1017/S0022112094002296} {\bibfield  {journal} {\bibinfo
  {journal} {Journal of Fluid Mechanics}\ }\textbf {\bibinfo {volume} {275}},\
  \bibinfo {pages} {83} (\bibinfo {year} {1994})}\BibitemShut {NoStop}%
\bibitem [{\citenamefont {Leonard}(1975)}]{Leonard1975237}%
  \BibitemOpen
  \bibfield  {author} {\bibinfo {author} {\bibfnamefont {A.}~\bibnamefont
  {Leonard}},\ }in\ \href {\doibase
  http://dx.doi.org/10.1016/S0065-2687(08)60464-1} {\emph {\bibinfo {booktitle}
  {Turbulent Diffusion in Environmental PollutionProceedings of a Symposium
  held at Charlottesville}}},\ \bibinfo {series} {Advances in Geophysics},
  Vol.\ \bibinfo {volume} {18, Part A},\ \bibinfo {editor} {edited by\ \bibinfo
  {editor} {\bibfnamefont {F.}~\bibnamefont {Frenkiel}}\ and\ \bibinfo {editor}
  {\bibfnamefont {R.}~\bibnamefont {Munn}}}\ (\bibinfo  {publisher}
  {Elsevier},\ \bibinfo {year} {1975})\ pp.\ \bibinfo {pages} {237 --
  248}\BibitemShut {NoStop}%
\bibitem [{\citenamefont {Yeo}(1987)}]{Yeo87}%
  \BibitemOpen
  \bibfield  {author} {\bibinfo {author} {\bibfnamefont {W.~K.}\ \bibnamefont
  {Yeo}},\ }\emph {\bibinfo {title} {{A generalized high pass/low pass
  averaging procedure for deriving and solving turbulent flow equations}}},\
  \href@noop {} {Ph.D. thesis},\ \bibinfo  {school} {The Ohio State University}
  (\bibinfo {year} {1987})\BibitemShut {NoStop}%
\bibitem [{\citenamefont {{Woodward}}\ \emph {et~al.}(2006)\citenamefont
  {{Woodward}}, \citenamefont {{Porter}}, \citenamefont {{Anderson}},
  \citenamefont {{Fuchs}},\ and\ \citenamefont {{Herwig}}}]{Woodward2006}%
  \BibitemOpen
  \bibfield  {author} {\bibinfo {author} {\bibfnamefont {P.~R.}\ \bibnamefont
  {{Woodward}}}, \bibinfo {author} {\bibfnamefont {D.~H.}\ \bibnamefont
  {{Porter}}}, \bibinfo {author} {\bibfnamefont {S.}~\bibnamefont
  {{Anderson}}}, \bibinfo {author} {\bibfnamefont {T.}~\bibnamefont {{Fuchs}}},
  \ and\ \bibinfo {author} {\bibfnamefont {F.}~\bibnamefont {{Herwig}}},\
  }\href {\doibase 10.1088/1742-6596/46/1/052} {\bibfield  {journal} {\bibinfo
  {journal} {Journal of Physics Conference Series}\ }\textbf {\bibinfo {volume}
  {46}},\ \bibinfo {pages} {370} (\bibinfo {year} {2006})}\BibitemShut
  {NoStop}%
\bibitem [{\citenamefont {Schmidt}\ and\ \citenamefont
  {Federrath}(2011)}]{Schmidt2011}%
  \BibitemOpen
  \bibfield  {author} {\bibinfo {author} {\bibfnamefont {W.}~\bibnamefont
  {Schmidt}}\ and\ \bibinfo {author} {\bibfnamefont {C.}~\bibnamefont
  {Federrath}},\ }\href {\doibase 10.1051/0004-6361/201015630} {\bibfield
  {journal} {\bibinfo  {journal} {Astron. Astrophys.}\ }\textbf {\bibinfo
  {volume} {528}},\ \bibinfo {pages} {A106} (\bibinfo {year}
  {2011})}\BibitemShut {NoStop}%
\bibitem [{\citenamefont {Bryan}\ \emph {et~al.}(2014)\citenamefont {Bryan},
  \citenamefont {Norman}, \citenamefont {O'Shea}, \citenamefont {Abel},
  \citenamefont {Wise}, \citenamefont {Turk}, \citenamefont {Reynolds},
  \citenamefont {Collins}, \citenamefont {Wang}, \citenamefont {Skillman},
  \citenamefont {Smith}, \citenamefont {Harkness}, \citenamefont {Bordner},
  \citenamefont {hoon Kim}, \citenamefont {Kuhlen}, \citenamefont {Xu},
  \citenamefont {Goldbaum}, \citenamefont {Hummels}, \citenamefont {Kritsuk},
  \citenamefont {Tasker}, \citenamefont {Skory}, \citenamefont {Simpson},
  \citenamefont {Hahn}, \citenamefont {Oishi}, \citenamefont {So},
  \citenamefont {Zhao}, \citenamefont {Cen}, \citenamefont {Li},\ and\
  \citenamefont {Collaboration}}]{Enzo2013}%
  \BibitemOpen
  \bibfield  {author} {\bibinfo {author} {\bibfnamefont {G.~L.}\ \bibnamefont
  {Bryan}}, \bibinfo {author} {\bibfnamefont {M.~L.}\ \bibnamefont {Norman}},
  \bibinfo {author} {\bibfnamefont {B.~W.}\ \bibnamefont {O'Shea}}, \bibinfo
  {author} {\bibfnamefont {T.}~\bibnamefont {Abel}}, \bibinfo {author}
  {\bibfnamefont {J.~H.}\ \bibnamefont {Wise}}, \bibinfo {author}
  {\bibfnamefont {M.~J.}\ \bibnamefont {Turk}}, \bibinfo {author}
  {\bibfnamefont {D.~R.}\ \bibnamefont {Reynolds}}, \bibinfo {author}
  {\bibfnamefont {D.~C.}\ \bibnamefont {Collins}}, \bibinfo {author}
  {\bibfnamefont {P.}~\bibnamefont {Wang}}, \bibinfo {author} {\bibfnamefont
  {S.~W.}\ \bibnamefont {Skillman}}, \bibinfo {author} {\bibfnamefont
  {B.}~\bibnamefont {Smith}}, \bibinfo {author} {\bibfnamefont {R.~P.}\
  \bibnamefont {Harkness}}, \bibinfo {author} {\bibfnamefont {J.}~\bibnamefont
  {Bordner}}, \bibinfo {author} {\bibfnamefont {J.}~\bibnamefont {hoon Kim}},
  \bibinfo {author} {\bibfnamefont {M.}~\bibnamefont {Kuhlen}}, \bibinfo
  {author} {\bibfnamefont {H.}~\bibnamefont {Xu}}, \bibinfo {author}
  {\bibfnamefont {N.}~\bibnamefont {Goldbaum}}, \bibinfo {author}
  {\bibfnamefont {C.}~\bibnamefont {Hummels}}, \bibinfo {author} {\bibfnamefont
  {A.~G.}\ \bibnamefont {Kritsuk}}, \bibinfo {author} {\bibfnamefont
  {E.}~\bibnamefont {Tasker}}, \bibinfo {author} {\bibfnamefont
  {S.}~\bibnamefont {Skory}}, \bibinfo {author} {\bibfnamefont {C.~M.}\
  \bibnamefont {Simpson}}, \bibinfo {author} {\bibfnamefont {O.}~\bibnamefont
  {Hahn}}, \bibinfo {author} {\bibfnamefont {J.~S.}\ \bibnamefont {Oishi}},
  \bibinfo {author} {\bibfnamefont {G.~C.}\ \bibnamefont {So}}, \bibinfo
  {author} {\bibfnamefont {F.}~\bibnamefont {Zhao}}, \bibinfo {author}
  {\bibfnamefont {R.}~\bibnamefont {Cen}}, \bibinfo {author} {\bibfnamefont
  {Y.}~\bibnamefont {Li}}, \ and\ \bibinfo {author} {\bibfnamefont {T.~E.}\
  \bibnamefont {Collaboration}},\ }\href
  {http://stacks.iop.org/0067-0049/211/i=2/a=19} {\bibfield  {journal}
  {\bibinfo  {journal} {The Astrophysical Journal Supplement Series}\ }\textbf
  {\bibinfo {volume} {211}},\ \bibinfo {pages} {19} (\bibinfo {year}
  {2014})}\BibitemShut {NoStop}%
\bibitem [{\citenamefont {Fryxell}\ \emph {et~al.}(2000)\citenamefont
  {Fryxell}, \citenamefont {Olson}, \citenamefont {Ricker}, \citenamefont
  {Timmes}, \citenamefont {Zingale}, \citenamefont {Lamb}, \citenamefont
  {MacNeice}, \citenamefont {Rosner}, \citenamefont {Truran},\ and\
  \citenamefont {Tufo}}]{Fryxell2000}%
  \BibitemOpen
  \bibfield  {author} {\bibinfo {author} {\bibfnamefont {B.}~\bibnamefont
  {Fryxell}}, \bibinfo {author} {\bibfnamefont {K.}~\bibnamefont {Olson}},
  \bibinfo {author} {\bibfnamefont {P.}~\bibnamefont {Ricker}}, \bibinfo
  {author} {\bibfnamefont {F.~X.}\ \bibnamefont {Timmes}}, \bibinfo {author}
  {\bibfnamefont {M.}~\bibnamefont {Zingale}}, \bibinfo {author} {\bibfnamefont
  {D.~Q.}\ \bibnamefont {Lamb}}, \bibinfo {author} {\bibfnamefont
  {P.}~\bibnamefont {MacNeice}}, \bibinfo {author} {\bibfnamefont
  {R.}~\bibnamefont {Rosner}}, \bibinfo {author} {\bibfnamefont {J.~W.}\
  \bibnamefont {Truran}}, \ and\ \bibinfo {author} {\bibfnamefont
  {H.}~\bibnamefont {Tufo}},\ }\href {\doibase 10.1086/317361} {\bibfield
  {journal} {\bibinfo  {journal} {Astrophys. J. Suppl. Ser.}\ }\textbf
  {\bibinfo {volume} {131}},\ \bibinfo {pages} {273} (\bibinfo {year}
  {2000})}\BibitemShut {NoStop}%
\bibitem [{\citenamefont {Pope}(2000)}]{Pope2000}%
  \BibitemOpen
  \bibfield  {author} {\bibinfo {author} {\bibfnamefont {S.~B.}\ \bibnamefont
  {Pope}},\ }\href {http://dx.doi.org/10.1017/CBO9780511840531} {\emph
  {\bibinfo {title} {Turbulent Flows}}}\ (\bibinfo  {publisher} {Cambridge
  University Press},\ \bibinfo {year} {2000})\ \bibinfo {note} {cambridge Books
  Online}\BibitemShut {NoStop}%
\bibitem [{\citenamefont {{Schmidt}}\ \emph {et~al.}(2009)\citenamefont
  {{Schmidt}}, \citenamefont {{Federrath}}, \citenamefont {{Hupp}},
  \citenamefont {{Kern}},\ and\ \citenamefont {{Niemeyer}}}]{Schmidt2009}%
  \BibitemOpen
  \bibfield  {author} {\bibinfo {author} {\bibfnamefont {W.}~\bibnamefont
  {{Schmidt}}}, \bibinfo {author} {\bibfnamefont {C.}~\bibnamefont
  {{Federrath}}}, \bibinfo {author} {\bibfnamefont {M.}~\bibnamefont {{Hupp}}},
  \bibinfo {author} {\bibfnamefont {S.}~\bibnamefont {{Kern}}}, \ and\ \bibinfo
  {author} {\bibfnamefont {J.~C.}\ \bibnamefont {{Niemeyer}}},\ }\href
  {\doibase 10.1051/0004-6361:200809967} {\bibfield  {journal} {\bibinfo
  {journal} {Astronomy \& Astrophysics}\ }\textbf {\bibinfo {volume} {494}},\
  \bibinfo {pages} {127} (\bibinfo {year} {2009})}\BibitemShut {NoStop}%
\bibitem [{\citenamefont {{Federrath}}\ \emph {et~al.}(2010)\citenamefont
  {{Federrath}}, \citenamefont {{Roman-Duval}}, \citenamefont {{Klessen}},
  \citenamefont {{Schmidt}},\ and\ \citenamefont {{Mac
  Low}}}]{2010A&A...512A..81F}%
  \BibitemOpen
  \bibfield  {author} {\bibinfo {author} {\bibfnamefont {C.}~\bibnamefont
  {{Federrath}}}, \bibinfo {author} {\bibfnamefont {J.}~\bibnamefont
  {{Roman-Duval}}}, \bibinfo {author} {\bibfnamefont {R.~S.}\ \bibnamefont
  {{Klessen}}}, \bibinfo {author} {\bibfnamefont {W.}~\bibnamefont
  {{Schmidt}}}, \ and\ \bibinfo {author} {\bibfnamefont {M.-M.}\ \bibnamefont
  {{Mac Low}}},\ }\href {\doibase 10.1051/0004-6361/200912437} {\bibfield
  {journal} {\bibinfo  {journal} {\aap}\ }\textbf {\bibinfo {volume} {512}},\
  \bibinfo {eid} {A81} (\bibinfo {year} {2010})}\BibitemShut {NoStop}%
\bibitem [{\citenamefont {Toro}(2009)}]{toro2009riemann}%
  \BibitemOpen
  \bibfield  {author} {\bibinfo {author} {\bibfnamefont {E.~F.}\ \bibnamefont
  {Toro}},\ }\href@noop {} {\emph {\bibinfo {title} {Riemann solvers and
  numerical methods for fluid dynamics: a practical introduction}}}\ (\bibinfo
  {publisher} {Springer Science \& Business Media},\ \bibinfo {year}
  {2009})\BibitemShut {NoStop}%
\bibitem [{\citenamefont {Miyoshi}\ and\ \citenamefont
  {Kusano}(2005)}]{Miyoshi2005315}%
  \BibitemOpen
  \bibfield  {author} {\bibinfo {author} {\bibfnamefont {T.}~\bibnamefont
  {Miyoshi}}\ and\ \bibinfo {author} {\bibfnamefont {K.}~\bibnamefont
  {Kusano}},\ }\href {\doibase http://dx.doi.org/10.1016/j.jcp.2005.02.017}
  {\bibfield  {journal} {\bibinfo  {journal} {Journal of Computational
  Physics}\ }\textbf {\bibinfo {volume} {208}},\ \bibinfo {pages} {315 }
  (\bibinfo {year} {2005})}\BibitemShut {NoStop}%
\bibitem [{\citenamefont {Federrath}\ \emph {et~al.}(2011)\citenamefont
  {Federrath}, \citenamefont {Chabrier}, \citenamefont {Schober}, \citenamefont
  {Banerjee}, \citenamefont {Klessen},\ and\ \citenamefont
  {Schleicher}}]{Federrath2011}%
  \BibitemOpen
  \bibfield  {author} {\bibinfo {author} {\bibfnamefont {C.}~\bibnamefont
  {Federrath}}, \bibinfo {author} {\bibfnamefont {G.}~\bibnamefont {Chabrier}},
  \bibinfo {author} {\bibfnamefont {J.}~\bibnamefont {Schober}}, \bibinfo
  {author} {\bibfnamefont {R.}~\bibnamefont {Banerjee}}, \bibinfo {author}
  {\bibfnamefont {R.~S.}\ \bibnamefont {Klessen}}, \ and\ \bibinfo {author}
  {\bibfnamefont {D.~R.~G.}\ \bibnamefont {Schleicher}},\ }\href {\doibase
  10.1103/PhysRevLett.107.114504} {\bibfield  {journal} {\bibinfo  {journal}
  {Phys. Rev. Lett.}\ }\textbf {\bibinfo {volume} {107}},\ \bibinfo {pages}
  {114504} (\bibinfo {year} {2011})}\BibitemShut {NoStop}%
\bibitem [{\citenamefont {Federrath}\ \emph {et~al.}(2014)\citenamefont
  {Federrath}, \citenamefont {Schober}, \citenamefont {Bovino},\ and\
  \citenamefont {Schleicher}}]{Federrath2014}%
  \BibitemOpen
  \bibfield  {author} {\bibinfo {author} {\bibfnamefont {C.}~\bibnamefont
  {Federrath}}, \bibinfo {author} {\bibfnamefont {J.}~\bibnamefont {Schober}},
  \bibinfo {author} {\bibfnamefont {S.}~\bibnamefont {Bovino}}, \ and\ \bibinfo
  {author} {\bibfnamefont {D.~R.~G.}\ \bibnamefont {Schleicher}},\ }\href
  {\doibase 10.1088/2041-8205/797/2/L19} {\bibfield  {journal} {\bibinfo
  {journal} {Astrophys. J.}\ }\textbf {\bibinfo {volume} {797}},\ \bibinfo
  {pages} {L19} (\bibinfo {year} {2014})}\BibitemShut {NoStop}%
\bibitem [{\citenamefont {{Waagan}}, \citenamefont {{Federrath}},\ and\
  \citenamefont {{Klingenberg}}(2011)}]{Waagan2011}%
  \BibitemOpen
  \bibfield  {author} {\bibinfo {author} {\bibfnamefont {K.}~\bibnamefont
  {{Waagan}}}, \bibinfo {author} {\bibfnamefont {C.}~\bibnamefont
  {{Federrath}}}, \ and\ \bibinfo {author} {\bibfnamefont {C.}~\bibnamefont
  {{Klingenberg}}},\ }\href {\doibase 10.1016/j.jcp.2011.01.026} {\bibfield
  {journal} {\bibinfo  {journal} {Journal of Computational Physics}\ }\textbf
  {\bibinfo {volume} {230}},\ \bibinfo {pages} {3331} (\bibinfo {year}
  {2011})}\BibitemShut {NoStop}%
\bibitem [{\citenamefont {Dedner}\ \emph {et~al.}(2002)\citenamefont {Dedner},
  \citenamefont {Kemm}, \citenamefont {Kröner}, \citenamefont {Munz},
  \citenamefont {Schnitzer},\ and\ \citenamefont {Wesenberg}}]{Dedner2002}%
  \BibitemOpen
  \bibfield  {author} {\bibinfo {author} {\bibfnamefont {A.}~\bibnamefont
  {Dedner}}, \bibinfo {author} {\bibfnamefont {F.}~\bibnamefont {Kemm}},
  \bibinfo {author} {\bibfnamefont {D.}~\bibnamefont {Kröner}}, \bibinfo
  {author} {\bibfnamefont {C.-D.}\ \bibnamefont {Munz}}, \bibinfo {author}
  {\bibfnamefont {T.}~\bibnamefont {Schnitzer}}, \ and\ \bibinfo {author}
  {\bibfnamefont {M.}~\bibnamefont {Wesenberg}},\ }\href {\doibase
  10.1006/jcph.2001.6961} {\bibfield  {journal} {\bibinfo  {journal} {Journal
  of Computational Physics}\ }\textbf {\bibinfo {volume} {175}},\ \bibinfo
  {pages} {645 } (\bibinfo {year} {2002})}\BibitemShut {NoStop}%
\bibitem [{\citenamefont {Kitsionas}\ \emph {et~al.}(2009)\citenamefont
  {Kitsionas}, \citenamefont {Federrath}, \citenamefont {Klessen},
  \citenamefont {Schmidt}, \citenamefont {Price}, \citenamefont {Dursi},
  \citenamefont {Gritschneder}, \citenamefont {Walch}, \citenamefont {Piontek},
  \citenamefont {Kim}, \citenamefont {Jappsen}, \citenamefont {Ciecielag},\
  and\ \citenamefont {{Mac Low}}}]{Kitsionas2009}%
  \BibitemOpen
  \bibfield  {author} {\bibinfo {author} {\bibfnamefont {S.}~\bibnamefont
  {Kitsionas}}, \bibinfo {author} {\bibfnamefont {C.}~\bibnamefont
  {Federrath}}, \bibinfo {author} {\bibfnamefont {R.~S.}\ \bibnamefont
  {Klessen}}, \bibinfo {author} {\bibfnamefont {W.}~\bibnamefont {Schmidt}},
  \bibinfo {author} {\bibfnamefont {D.~J.}\ \bibnamefont {Price}}, \bibinfo
  {author} {\bibfnamefont {L.~J.}\ \bibnamefont {Dursi}}, \bibinfo {author}
  {\bibfnamefont {M.}~\bibnamefont {Gritschneder}}, \bibinfo {author}
  {\bibfnamefont {S.}~\bibnamefont {Walch}}, \bibinfo {author} {\bibfnamefont
  {R.}~\bibnamefont {Piontek}}, \bibinfo {author} {\bibfnamefont
  {J.}~\bibnamefont {Kim}}, \bibinfo {author} {\bibfnamefont {A.-K.}\
  \bibnamefont {Jappsen}}, \bibinfo {author} {\bibfnamefont {P.}~\bibnamefont
  {Ciecielag}}, \ and\ \bibinfo {author} {\bibfnamefont {M.-M.}\ \bibnamefont
  {{Mac Low}}},\ }\href {\doibase 10.1051/0004-6361/200811170} {\bibfield
  {journal} {\bibinfo  {journal} {Astron. Astrophys.}\ }\textbf {\bibinfo
  {volume} {508}},\ \bibinfo {pages} {541} (\bibinfo {year}
  {2009})}\BibitemShut {NoStop}%
\bibitem [{\citenamefont {Newville}\ \emph {et~al.}(2014)\citenamefont
  {Newville}, \citenamefont {Stensitzki}, \citenamefont {Allen},\ and\
  \citenamefont {Ingargiola}}]{newville_2014_11813}%
  \BibitemOpen
  \bibfield  {author} {\bibinfo {author} {\bibfnamefont {M.}~\bibnamefont
  {Newville}}, \bibinfo {author} {\bibfnamefont {T.}~\bibnamefont
  {Stensitzki}}, \bibinfo {author} {\bibfnamefont {D.~B.}\ \bibnamefont
  {Allen}}, \ and\ \bibinfo {author} {\bibfnamefont {A.}~\bibnamefont
  {Ingargiola}},\ }\href {\doibase 10.5281/zenodo.11813} {\enquote {\bibinfo
  {title} {{LMFIT: Non-Linear Least-Square Minimization and Curve-Fitting for
  Python}},}\ } (\bibinfo {year} {2014})\BibitemShut {NoStop}%
\bibitem [{\citenamefont {Dallas}\ and\ \citenamefont
  {Alexakis}(2013)}]{Dallas2013}%
  \BibitemOpen
  \bibfield  {author} {\bibinfo {author} {\bibfnamefont {V.}~\bibnamefont
  {Dallas}}\ and\ \bibinfo {author} {\bibfnamefont {A.}~\bibnamefont
  {Alexakis}},\ }\href {\doibase 10.1063/1.4824195} {\bibfield  {journal}
  {\bibinfo  {journal} {Phys. Fluids}\ }\textbf {\bibinfo {volume} {25}},\
  \bibinfo {pages} {105106} (\bibinfo {year} {2013})},\ \Eprint
  {http://arxiv.org/abs/1304.0695} {arXiv:1304.0695} \BibitemShut {NoStop}%
\bibitem [{sup()}]{suppmat}%
  \BibitemOpen
  \href@noop {} {}\bibinfo {note} {See supplemental material at [URL will be
  inserted by AIP] for detailed correlation and coefficient values split by
  simulation.}\BibitemShut {Stop}%
\bibitem [{\citenamefont {{Mac Low}}\ and\ \citenamefont
  {{Klessen}}(2004)}]{2004RvMP...76..125M}%
  \BibitemOpen
  \bibfield  {author} {\bibinfo {author} {\bibfnamefont {M.-M.}\ \bibnamefont
  {{Mac Low}}}\ and\ \bibinfo {author} {\bibfnamefont {R.~S.}\ \bibnamefont
  {{Klessen}}},\ }\href {\doibase 10.1103/RevModPhys.76.125} {\bibfield
  {journal} {\bibinfo  {journal} {Reviews of Modern Physics}\ }\textbf
  {\bibinfo {volume} {76}},\ \bibinfo {pages} {125} (\bibinfo {year}
  {2004})}\BibitemShut {NoStop}%
\bibitem [{\citenamefont {Pakmor}, \citenamefont {Marinacci},\ and\
  \citenamefont {Springel}(2014)}]{2041-8205-783-1-L20}%
  \BibitemOpen
  \bibfield  {author} {\bibinfo {author} {\bibfnamefont {R.}~\bibnamefont
  {Pakmor}}, \bibinfo {author} {\bibfnamefont {F.}~\bibnamefont {Marinacci}}, \
  and\ \bibinfo {author} {\bibfnamefont {V.}~\bibnamefont {Springel}},\ }\href
  {http://stacks.iop.org/2041-8205/783/i=1/a=L20} {\bibfield  {journal}
  {\bibinfo  {journal} {The Astrophysical Journal Letters}\ }\textbf {\bibinfo
  {volume} {783}},\ \bibinfo {pages} {L20} (\bibinfo {year}
  {2014})}\BibitemShut {NoStop}%
\end{thebibliography}
\end{document}